\documentclass[letter,longauth]{aa}
\usepackage{rotating}
\usepackage{placeins}
\usepackage{gensymb}
\usepackage{subfigure}
\usepackage{orcidlink}
\usepackage{graphicx}
\usepackage{txfonts}
\usepackage[]{hyperref}
\usepackage{newunicodechar}
\DeclareRobustCommand{\okina}{%
  \raisebox{\dimexpr\fontcharht\font`A-\height}{%
    \scalebox{0.8}{`}%
  }%
}
\newunicodechar{ʻ}{\okina}

\begin{document}

   \title{Expanding the frontiers of cool-dwarf asteroseismology\\ with ESPRESSO}

   \subtitle{Detection of solar-like oscillations in the K5 dwarf $\epsilon$ Indi}

   \titlerunning{Expanding the frontiers of cool-dwarf asteroseismology}

   \authorrunning{T. L. Campante et al.}

   \author{
        T. L. Campante\inst{\ref{IA}}\fnmsep\inst{\ref{FCUP}}\orcidlink{0000-0002-4588-5389}
    \and
        H. Kjeldsen\inst{\ref{Aarhus}}\orcidlink{0000-0002-9037-0018}
    \and
        Y. Li\inst{\ref{Hawaii}}\orcidlink{0000-0003-3020-4437}
    \and
        M. N. Lund\inst{\ref{Aarhus}}\orcidlink{0000-0001-9214-5642}
    \and
        A. M. Silva\inst{\ref{IA}}\fnmsep\inst{\ref{FCUP}}\orcidlink{0000-0003-4920-738X}
    \and
        E. Corsaro\inst{\ref{Catania}}\orcidlink{0000-0001-8835-2075}
    \and
        J. Gomes da Silva\inst{\ref{IA}}\orcidlink{0000-0001-8056-9202}
    \and
        J. H. C. Martins\inst{\ref{IA}}\orcidlink{0000-0002-1532-9082}
    \and
        V. Adibekyan\inst{\ref{IA}}\fnmsep\inst{\ref{FCUP}}\orcidlink{0000-0002-0601-6199}
    \and
        T. Azevedo Silva\inst{\ref{IA}}\fnmsep\inst{\ref{FCUP}}\orcidlink{0000-0002-9379-4895}
    \and
        T. R. Bedding\inst{\ref{Sydney}}\orcidlink{0000-0001-5222-4661}
    \and
        D. Bossini\inst{\ref{IA}}\orcidlink{0000-0002-9480-8400}
    \and
        D. L. Buzasi\inst{\ref{Florida}}\orcidlink{0000-0002-1988-143X}
    \and
        W. J. Chaplin\inst{\ref{Bham}}\orcidlink{0000-0002-5714-8618}
    \and
        R. R. Costa\inst{\ref{IA}}\fnmsep\inst{\ref{FCUP}}\orcidlink{0009-0008-6039-6381}
    \and
        M. S. Cunha\inst{\ref{IA}}\orcidlink{0000-0001-8237-7343}
    \and
        E. Cristo\inst{\ref{IA}}\fnmsep\inst{\ref{FCUP}}\orcidlink{0000-0001-5992-7589}
    \and
        J. P. Faria\inst{\ref{IA}}\fnmsep\inst{\ref{FCUP}}\orcidlink{0000-0002-6728-244X}
    \and
        R. A. Garc\'ia\inst{\ref{CEA}}\orcidlink{0000-0002-8854-3776}
    \and
        D. Huber\inst{\ref{Hawaii}}\orcidlink{0000-0001-8832-4488}
    \and
        M. S. Lundkvist\inst{\ref{Aarhus}}\orcidlink{0000-0002-8661-2571}
    \and
        T. S. Metcalfe\inst{\ref{WDRC}}\orcidlink{0000-0003-4034-0416}
    \and
        M. J. P. F. G. Monteiro\inst{\ref{IA}}\fnmsep\inst{\ref{FCUP}}\orcidlink{0000-0003-0513-8116}
    \and
        A. W. Neitzel\inst{\ref{IA}}\fnmsep\inst{\ref{FCUP}}\orcidlink{0000-0001-6283-907X}
    \and
        M. B. Nielsen\inst{\ref{Bham}}\orcidlink{0000-0001-9169-2599}
    \and
        E. Poretti\inst{\ref{Brera}}\orcidlink{0000-0003-1200-0473}
    \and
        N. C. Santos\inst{\ref{IA}}\fnmsep\inst{\ref{FCUP}}\orcidlink{0000-0003-4422-2919}
    \and
        S. G. Sousa\inst{\ref{IA}}\orcidlink{0000-0001-9047-2965}
   }

   \institute{
        Instituto de Astrof\'{\i}sica e Ci\^{e}ncias do Espa\c{c}o, Universidade do Porto,  Rua das Estrelas, 4150-762 Porto, Portugal\\
        \email{tiago.campante@astro.up.pt}
        \label{IA}
   \and
        Departamento de F\'{\i}sica e Astronomia, Faculdade de Ci\^{e}ncias da Universidade do Porto, Rua do Campo Alegre, s/n, 4169-007 Porto, Portugal
        \label{FCUP}
    \and
        Stellar Astrophysics Centre (SAC), Department of Physics and Astronomy, Aarhus University, Ny Munkegade 120, 8000 Aarhus C, Denmark
        \label{Aarhus}
    \and
        Institute for Astronomy, University of Hawai\okina i, 2680 Woodlawn Drive, Honolulu, HI 96822, USA
        \label{Hawaii}
    \and
        INAF --- Osservatorio Astrofisico di Catania, Via S.~Sofia 78, 95123 Catania, Italy
        \label{Catania}
    \and
        Sydney Institute for Astronomy (SIfA), School of Physics, University of Sydney, NSW 2006, Australia
        \label{Sydney}
    \and
        Department of Chemistry and Physics, Florida Gulf Coast University, 10501 FGCU Blvd.~S., Fort Myers, FL 33965, USA
        \label{Florida}
    \and
        School of Physics and Astronomy, University of Birmingham, Edgbaston, Birmingham B15 2TT, UK
        \label{Bham}
    \and
        Universit\'e Paris-Saclay, Universit\'e Paris Cit\'e, CEA, CNRS, AIM, 91191, Gif-sur-Yvette, France
        \label{CEA}
    \and
        White Dwarf Research Corporation, 9020 Brumm Trail, Golden, CO 80403, USA
        \label{WDRC}
    \and
        INAF --- Osservatorio Astronomico di Brera, Via E.~Bianchi 46, 23807 Merate, Italy
        \label{Brera}
   }

   \date{Received Month Day, Year; accepted Month Day, Year}
 
   \abstract  
   {Fuelled by space photometry, asteroseismology is vastly benefitting the study of cool main-sequence stars, which exhibit convection-driven solar-like oscillations. Even so, the tiny oscillation amplitudes in K dwarfs continue to pose a challenge to space-based asteroseismology. A viable alternative is offered by the lower stellar noise over the oscillation timescales in Doppler observations. In this letter we present the definite detection of solar-like oscillations in the bright K5 dwarf $\epsilon$~Indi based on time-intensive observations collected with the ESPRESSO spectrograph at the VLT, thus making it the coolest seismic dwarf ever observed. We measured the frequencies of a total of 19 modes of degree $\ell\!=\!0$--2 along with $\nu_{\rm max}\!=\!5305\pm176\:{\rm \mu Hz}$ and $\Delta\nu\!=\!201.25\pm0.16\:{\rm \mu Hz}$. The peak amplitude of radial modes is $2.6\pm0.5\:{\rm cm\,s^{-1}}$, or a mere ${\sim} 14\%$ of the solar value. Measured mode amplitudes are ${\sim} 2$ times lower than predicted from a nominal $L/M$ scaling relation and favour a scaling closer to $(L/M)^{1.5}$ below ${\sim} 5500\:{\rm K}$, carrying important implications for our understanding of the coupling efficiency between pulsations and near-surface convection in K dwarfs. This detection conclusively shows that precise asteroseismology of cool dwarfs is possible down to at least the mid-K regime using next-generation spectrographs on large-aperture telescopes, effectively opening up a new domain in observational asteroseismology.}

   \keywords{Asteroseismology -- Stars: individual: $\epsilon$~Indi~A -- Stars: late-type -- Stars: oscillations (including pulsations) -- Techniques: radial velocities}

   \maketitle

\nolinenumbers
\section{Introduction}\label{sec:Intro}

Asteroseismology has seen remarkable advances thanks to missions such as Convection, Rotation and planetary Transits \citep[CoRoT;][]{CoRoT} and Kepler/K2 \citep[][]{Kepler,K2}. These missions have provided exquisite space photometry, enabling the detailed study of the interiors of solar-type and red-giant stars, which exhibit convection-driven solar-like oscillations \citep[for a recent review, see][]{Aerts21}. The ongoing Transiting Exoplanet Survey Satellite \citep[TESS;][]{TESS}, along with the upcoming PLAnetary Transits and Oscillations of stars \citep[PLATO;][]{PLATO} and Nancy Grace Roman \citep[][]{Roman} space telescopes, are set to revolutionise the field as they are expected to raise the yield of known solar-like oscillators to a few million stars, or by two orders of magnitude over previous missions combined \citep[][]{Gould15,Miglio17,Hon21,Goupil24}.

\begin{figure}[!t]
\centering
\includegraphics[width=0.5\textwidth,trim=0.4cm 0.5cm 0cm 0.275cm,clip]{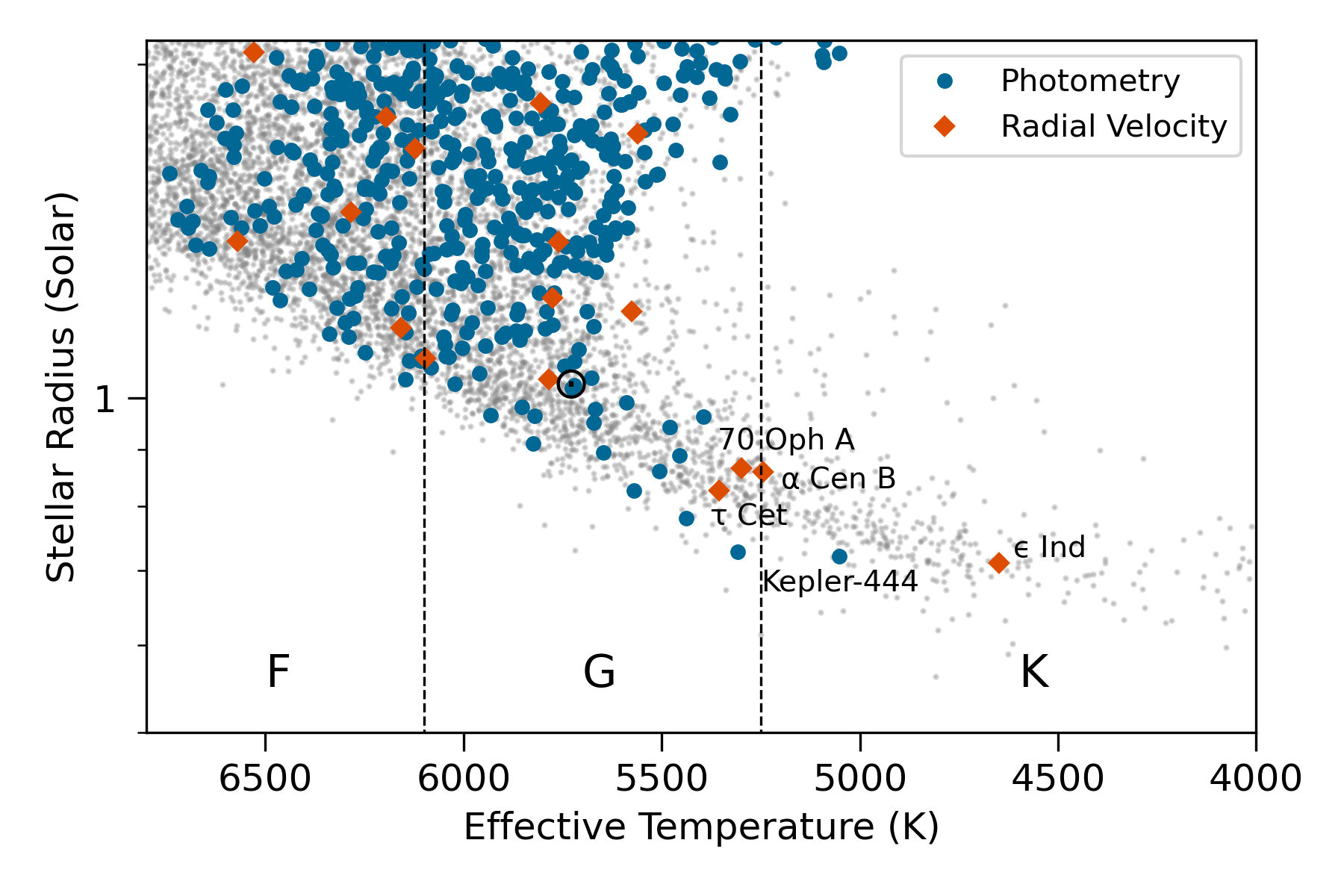}
\caption{Stellar radius-effective temperature diagram highlighting seismic detections from Kepler and TESS photometry \citep[blue circles;][]{Mathur17,Hatt23}, and radial-velocity campaigns \citep[red diamonds; see e.g.][and references therein]{Arentoft08,Kjeldsen08}. The stellar background sample (grey dots) is taken from the TESS Input Catalog \citep[TIC;][]{TIC}. The Sun is represented by its usual symbol. Approximate spectral type ranges (F, G, and K) are delimited by the vertical dashed lines. $\epsilon$~Indi (K5~V) is the coolest seismic dwarf observed to date \citep[its interferometric radius and effective temperature were used to place it in the diagram;][]{Rains20}.}
\label{fig:HRD}
\end{figure}

Despite this success story, space-based asteroseismology faces a challenge regarding K dwarfs. Owing to the low luminosities of K dwarfs, their oscillation amplitudes are extremely small \citep[below a few parts per million or, equivalently, $10\:{\rm cm\,s^{-1}}$;][]{Kjeldsen08,Verner11,Corsaro13} and thus hard to detect, even with multi-year Kepler photometry  \citep[e.g. Kepler-444;][]{Kepler444}. As a result, only a few dwarfs cooler than the Sun have detected solar-like oscillations to date, and none cooler than ${\sim} 5000\:{\rm K}$ (see Fig.~\ref{fig:HRD}).

A viable alternative to space photometry is offered by Doppler observations. Stellar noise due to non-oscillatory fluctuations associated with activity and granulation is substantially lower in Doppler than it is in photometry \citep[][]{Harvey88}. Consequently, radial-velocity (RV) observations have a higher signal-to-noise ratio ($S/N$) over the typical timescales of the oscillations \citep[by an order of magnitude in power for the Sun;][]{Grundahl07}. This motivated a number of pre-Kepler ground-based campaigns on cool dwarfs with the then state-of-the-art spectrographs such as the High Accuracy Radial velocity Planet Searcher \citep[HARPS;][]{HARPS} and the Ultraviolet and Visual Echelle Spectrograph \citep[UVES;][]{UVES}. Observing runs like those on $\tau$~Ceti \citep[G8~V;][]{Teixeira09}, 70~Ophiuchi~A \citep[K0~V;][]{CarrierEggenberger06}, and $\alpha$~Centauri~B \citep[K1~V;][]{Kjeldsen05} are the epitome of such efforts, having helped set a lower effective temperature ($T_{\rm eff}$) bound on cool-dwarf asteroseismology. However, long readout times and/or relatively small apertures meant that these early campaigns would remain limited to the very brightest dwarfs.

K dwarfs have since become a primary focus in searches for potentially habitable planets \citep[][]{KOBE,HWO}. Moreover, owing to their ubiquity and long lives, they are unique probes of local Galactic chemical evolution \citep[][]{Adibekyan12,DelgadoMena21}. The time is thus ripe to   systematically  extend  asteroseismology to these cooler dwarfs via ultra-high-precision RV observations that make use of next-generation spectrographs on large-aperture telescopes. The combination of a large collecting area, instrumental stability, and high spectral resolution makes the Echelle SPectrograph for Rocky Exoplanets and Stable Spectroscopic Observations \citep[ESPRESSO;][]{ESPRESSO}, mounted on the Very Large Telescope (VLT) at the European Southern Observatory (ESO), Paranal, Chile, particularly suitable for this purpose.

An exploratory campaign on the seventh magnitude K3 dwarf \object{HD~40307} was conducted in December 2018 as part of the ESPRESSO Guaranteed Time Observations (GTO), as described in Sect.~5.3.2 of \citet{ESPRESSO}. However, due in part to the target's relative faintness, only a tentative claim of p-mode detection (at the level of 3--$4\:{\rm cm\,s^{-1}}$) could be made. In this letter we  overcome this drawback as we report on the recent campaign conducted with ESPRESSO on the fourth magnitude K5 dwarf \object{$\epsilon$~Indi~A} (HD~209100, HR~8387; hereafter $\epsilon$~Indi), a target providing a nearly ten times greater flux than HD~40307. We are able to firmly establish the presence of solar-like oscillations in the RV data of $\epsilon$~Indi, thus making it the coolest seismic dwarf observed to date.

\section{Observations and data reduction}\label{sec:Obs}

$\epsilon$~Indi is a nearby ($d\!=\!3.64\:{\rm pc}$), bright ($V\!=\!4.69$), and metal-poor \citep[${\rm [Fe/H]}\!=\!-0.17\pm0.05\:{\rm dex}$;][]{GomesdaSilva21} K5 dwarf \citep[interferometry-based $T_{\rm eff}\!=\!4649\pm84\:{\rm K}$;][]{Rains20}. It hosts a cold Jupiter ($\epsilon$~Indi~Ab) on a 45 yr period orbit detected in RV and astrometric data \citep[][]{Feng19}. $\epsilon$~Indi further hosts a brown dwarf binary ($\epsilon$~Indi~Ba, Bb) in a wide orbit with a projected separation of ${\sim} 1500\:{\rm AU}$ \citep[][]{McCaughrean04}. This system hence provides a benchmark for the study of the formation of gas-giant planets and brown dwarfs \citep[e.g.][]{Pathak21,Viswanath21,Chen22,Subjak23}.

We observed $\epsilon$~Indi for six consecutive  half nights  with ESPRESSO in September 2022. Observations were carried out in single Unit Telescope (single-UT) high-resolution ($1\times1$ binning and fast readout) mode. Weather conditions were generally favourable, with photometric and/or clear skies over the first two nights, and spells of thin cirrus clouds and relatively high winds during the remaining nights. We obtained 2084 spectra with a fixed exposure time of $25\:{\rm s}$ and a median cadence of one exposure every $60\:{\rm s}$ (which corresponds to a Nyquist frequency of $8.3\:{\rm mHz}$). The spectra were subsequently reduced using version 3.0.0 of the ESPRESSO data reduction software (DRS), having adopted a K6 stellar binary mask to compute cross-correlation functions (CCFs), from which RVs and associated CCF parameters (see Appendix~\ref{sec:RVs}) were derived.

\begin{figure}[!t]
\centering
\includegraphics[width=0.5\textwidth,trim=1.6cm 0.7cm 0.2cm 0.9cm,clip]{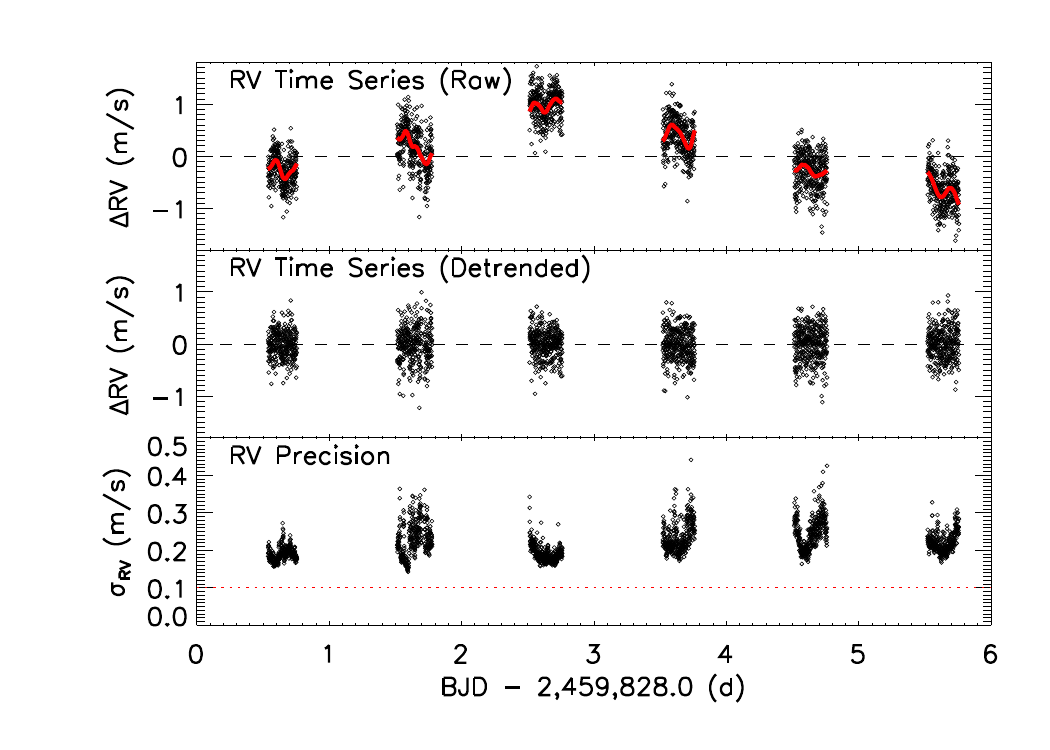}
\caption{Time series of ESPRESSO radial-velocity measurements of $\epsilon$~Indi. \emph{Top:} Raw time series (after removal of a constant RV offset). The solid red curves represent smoothing functions applied on a nightly basis (see text for details). \emph{Middle:} Detrended time series (after high-pass filtering). \emph{Bottom:} Internal (photon-noise limited) RV precision as returned by the ESPRESSO DRS. The horizontal dotted line represents the instrumental noise level of $10\:{\rm cm\,s^{-1}}$ quoted by \citet{ESPRESSO}.}
\label{fig:ts}
\end{figure}

The resulting (raw) RVs are shown in the top panel of Fig.~\ref{fig:ts}. The slow modulation of the time series is likely a manifestation, as seen in this short-duration data set, of the 18-day signal (corresponding to half the rotation period) due to rotational activity variations identified by \citet{Feng19}. Moreover, the time series displays intranight trends, presumably due to a combination of instrumental drift and stellar convection. To remove the slow modulation and intranight trends, we high-pass filtered the time series one night at a time using a triangular smoothing function (${\sim} 2$ hr cutoff). The detrended time series thus obtained is shown in the middle panel of Fig.~\ref{fig:ts}. Its dispersion (rms scatter) is $30\:{\rm cm\,s^{-1}}$ and greater than the photon noise, which we attribute to the presence of oscillations (the average photon-noise uncertainty per data point is $<\!\sigma_{\rm RV}\!>=22\:{\rm cm\,s^{-1}}$; see bottom panel of Fig.~\ref{fig:ts}).

\section{Asteroseismic data analysis}\label{sec:Seismic}

\subsection{Computation of the power spectrum}\label{sec:Pow}

We based our analysis of the power spectrum on the discrete Fourier transform (DFT) of the detrended RV time series. This involved using the measurement uncertainties, $\sigma_{{\rm RV},i}$, as statistical weights in calculating the power spectrum (according to $w_i=1/\sigma_{{\rm RV},i}^2$). In order to optimise the noise floor in the power spectrum, these weights were further adjusted to account for a small fraction of bad data points (37 data points, or ${\sim} 2\%$ of the total, were removed) as well as night-to-night variations in the noise level. We followed a well-tested procedure in adjusting the weights \citep[for details, see e.g.][]{Bedding04,Arentoft08}.

\begin{figure}[!t]
\centering
\includegraphics[width=0.5\textwidth,trim=1.0cm 0.25cm 0.1cm 0.8cm,clip]{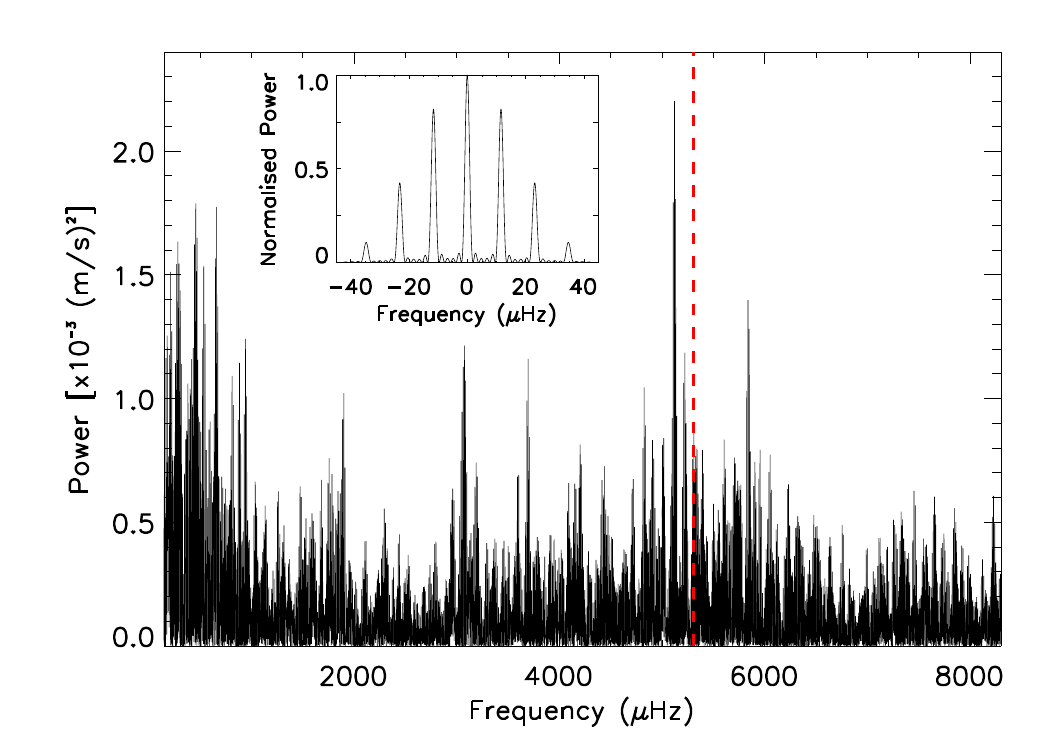}
\caption{Noise-optimised power spectrum of $\epsilon$~Indi. The power spectrum has been oversampled for visual purposes. A clear power excess due to solar-like oscillations can be seen centred just above $5\:{\rm mHz}$. The vertical dashed line represents the measured $\nu_{\rm max}$ (see text for details). The inset shows the spectral window, with prominent sidelobes (daily aliases) due to the single-site nature of the observations.}
\label{fig:ps}
\end{figure}

The resulting noise-optimised power spectrum is shown in Fig.~\ref{fig:ps}, which displays a clear power excess due to solar-like oscillations centred just above $5\:{\rm mHz}$ (typical periods of ${\sim} 3$ min). This is in agreement with the predicted frequency of maximum oscillation amplitude, $\nu_{\rm max}\,{=}\,\nu_{{\rm max},\sun}\,(g/g_\sun)\,(T_{\rm eff}/T_{{\rm eff},\sun})^{-1/2}\,{\sim}\,5.2\:{\rm mHz}$, scaled by solar values \citep[][]{Brown91,KjeldsenBedding95}, where $g$ is the surface gravity \citep[$\log g\!=\!4.61\pm0.29\:{\rm dex}$;][]{GomesdaSilva21}. We proceeded to measure $\nu_{\rm max}$ based on a heavily smoothed version of the power spectrum (see Sect.~\ref{sec:Amp} for details). After correcting for the background noise in the power spectrum, an estimate $\nu_{\rm max}\!=\!5305\pm176\:{\rm \mu Hz}$ was obtained.

The spectral window is shown as an inset in Fig.~\ref{fig:ps}, and reveals prominent sidelobes caused by the daily gaps in the RV data. The average photon-noise level in the amplitude spectrum, as measured at high frequencies (above $6.5\:{\rm mHz}$, i.e. beyond the frequency range occupied by the p modes), is $0.94\:{\rm cm\,s^{-1}}$. For comparison, the high-frequency noise level reported for the observations of $\alpha$~Centauri~B is $1.39\:{\rm cm\,s^{-1}}$ \citep[][]{Kjeldsen05}.

\subsection{Oscillation frequencies}\label{sec:Osc}

The frequencies of acoustic (p) modes of high radial order, $n$, and low angular degree, $\ell$, are well approximated by the asymptotic relation \citep[][]{Tassoul80}:
\begin{equation}
\nu_{n,\ell} \simeq \Delta\nu \left(n+\frac{\ell}{2}+\varepsilon\right) - \delta\nu_{0\ell} \, .
\label{eq:asymptotic}
\end{equation}
Here $\Delta\nu$ is the large separation between modes of like degree and consecutive order, being a probe of the mean stellar density; $\delta\nu_{0\ell}$ is the small separation between modes of different degree and is sensitive to variations in the sound speed gradient near the core in main-sequence stars; and the dimensionless offset, $\varepsilon$, is determined by the reflection properties of the surface layers. Observed solar-like oscillations in main-sequence stars are expected to follow this relation closely. We  thus used this prior information to guide the mode identification and extraction (as described below), bearing in mind the presence of daily aliases in the power spectrum (appearing at splittings of $\pm11.57\:{\rm \mu Hz}$, or $\pm1$ cycle per day, about genuine peaks).

\begingroup
\renewcommand{\arraystretch}{1.1} 
\begin{table}
\caption{Oscillation frequencies for $\epsilon$~Indi (in $\mu$Hz)}
\label{tab:freqs}
\centering
\begin{tabular}{c c c c}
\hline\hline
$n$ & $\ell\!=\!0$ & $\ell\!=\!1$ & $\ell\!=\!2$ \\
\hline
21 & $4518.59\pm1.15$ & $4618.27\pm1.35$ & $4703.16\pm1.33$ \\
22 & $4720.55\pm0.92$ & $4815.16\pm0.92$ & $4906.18\pm1.26$ \\
23 & $4919.93\pm0.87$ & $5017.56\pm0.83$ & $5107.26\pm0.98$ \\
24 & $5121.61\pm0.51$ & $5223.40\pm0.99$ & $5308.01\pm0.92$ \\
25 & $5322.46\pm0.91$ & $5416.88\pm1.34$ & $5509.48\pm1.31$ \\
26 & $5525.24\pm1.28$ & $5616.77\pm1.15$ & $\cdots$ \\
27 & $5726.30\pm1.13$ & $5826.26\pm1.03$ & $\cdots$ \\
\hline
\end{tabular}
\tablefoot{Quoted uncertainties depend on the $S/N$ of the corresponding mode peaks, and were calibrated using simulations \citep[e.g.][]{Kjeldsen05}. We opted to list mode frequencies without correcting for the line-of-sight motion \citep{Davies14}. Given the non-negligible magnitude of this effect (${\sim} 0.7\:{\rm \mu Hz}$ at $5000\:{\rm \mu Hz}$), we advise   applying this correction when directly comparing the observed individual frequencies to model frequencies.}
\end{table}
\endgroup

\begin{figure}[!t]
\centering
\includegraphics[width=0.48\textwidth,trim=0.1cm 0cm 0cm 0cm,clip]{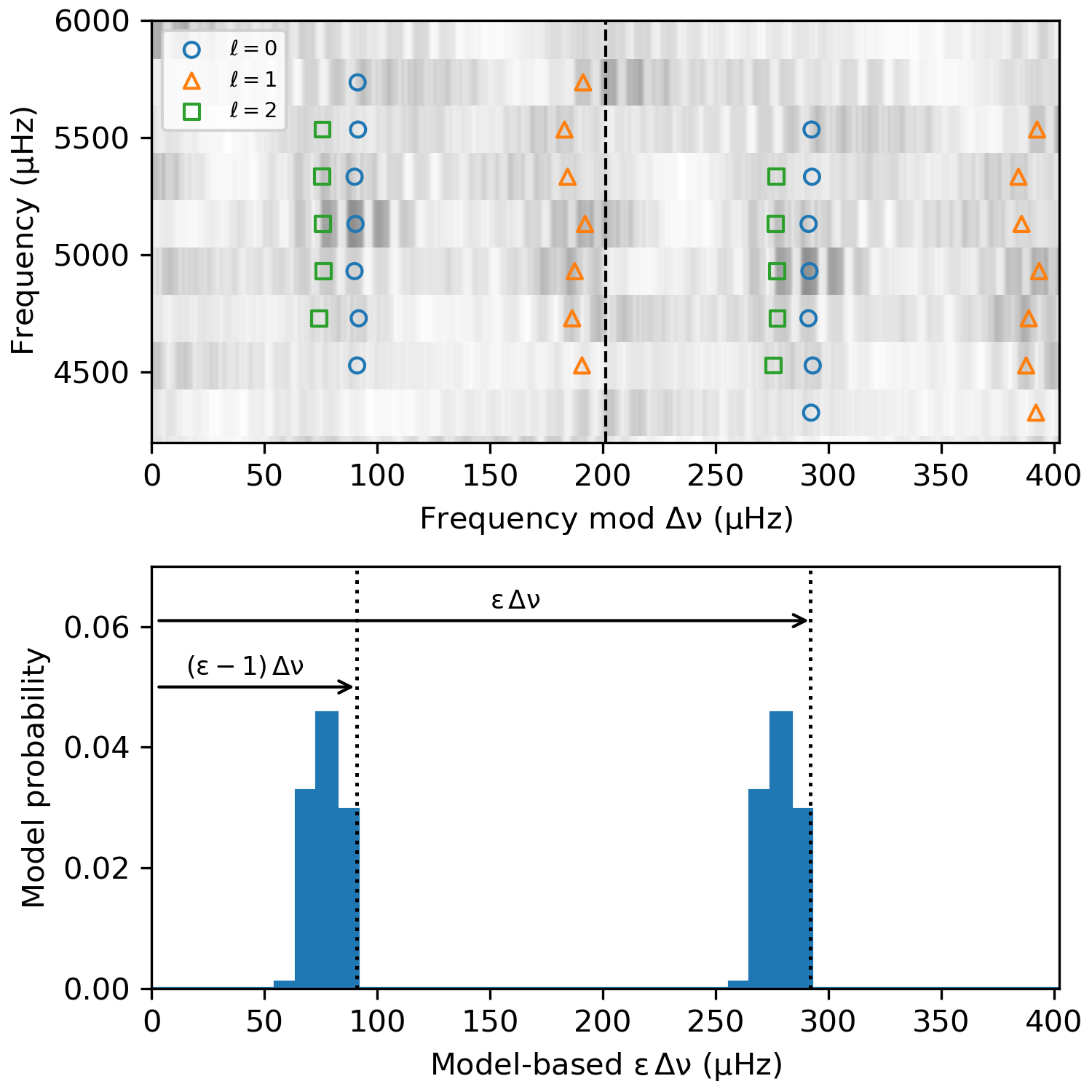}
\caption{Outcome of the mode identification and extraction procedure. \emph{Top:} Replicated \'echelle diagram displaying the mode frequencies extracted for $\epsilon$~Indi (cf.~Table~\ref{tab:freqs}). The smoothed power spectrum is shown in grey scale. The vertical dashed line gives the measured $\Delta\nu$ value. \emph{Bottom:} Likelihood-weighted histogram of $\varepsilon\,\Delta\nu$ for the models in the grid (see text for details). The observed $\varepsilon\,\Delta\nu$ is represented by a vertical dotted line.}
\label{fig:echelle}
\end{figure}

Extracted mode frequencies are listed in Table~\ref{tab:freqs} and displayed in \'echelle format in the top panel of Fig.~\ref{fig:echelle}. Owing to the short duration of the RV time series, individual modes are only partially resolved (see also Sect.~\ref{sec:Lifetimes}). Modes were thus extracted using a standard iterative sine-wave fitting procedure, also known as prewhitening \citep[e.g.][]{Bedding10}. A total of 19 modes of degree $\ell\!=\!0$--2 were extracted across seven orders down to $S/N\!=\!2.5$. The full procedure for identifying and extracting oscillation frequencies consisted in the following steps:
\begin{enumerate}
\item We measured the strongest peak ($5121.61\:{\rm \mu Hz}$) within the frequency range occupied by the p modes and used it to compute a modified comb response \citep[e.g.][]{Kjeldsen95} over a range of trial large separations. The comb response peaks at ${\sim} 201.3\:{\rm \mu Hz}$, which we adopted as a first estimate of $\Delta\nu$.

\item Guided by this estimate, we identified the sequence of (nearly) regularly spaced peaks below and above the dominant mode at $5121.61\:{\rm \mu Hz}$ (and hence sharing the same $\ell$). We measured seven such modes, based on which the large separation, $\Delta\nu\!=\!201.25\pm0.16\:{\rm \mu Hz}$, was computed. Based on the value for $\Delta\nu$ and the frequency of the dominant mode, we inferred $\varepsilon\!=\!1.451\pm0.019$, consistent with the  empirical results in the literature obtained for other cool dwarfs \citep[cf.][]{White11_1,White11_2,White12,Lund17}, and hence with these being radial ($\ell\!=\!0$) modes.

\item We extracted these modes from the time series through iterative sine-wave fitting. By collapsing the resulting prewhitened power spectrum about the position of the $\ell\!=\!0$ ridge, we saw a power excess at lower frequencies, which we assigned to the $\ell\!=\!2$ ridge, being able to resolve the small separation ($\delta\nu_{02}\!\sim\!14.5\:{\rm \mu Hz}$). Guided by this, we then identified a sequence of five quadrupole ($\ell\!=\!2$) modes in the prewhitened power spectrum, based on which we obtained $\delta\nu_{02}\!=\!15.28\pm0.45\:{\rm \mu Hz}$. We note that $\delta\nu_{02}$ is a decreasing function of frequency (or $n$), as expected \citep[e.g.][]{Lund17}.

\item Finally, we collapsed the power spectrum about the midpoint between consecutive radial modes. A clear power excess corresponding to the $\ell\!=\!1$ ridge could be seen below the midpoint frequency ($\delta\nu_{01}\!\sim\!4.4\:{\rm \mu Hz}$). Guided by this, we identified a sequence of seven dipole ($\ell\!=\!1$) modes, providing a direct measurement of the small separation, $\delta\nu_{01}\!=\!3.46\pm1.48\:{\rm \mu Hz}$. We estimated the power ratio between the $\ell\!=\!1$ and 0 ridges to be ${\sim} 1.3$, in accordance with the predicted spatial response of Doppler observations \citep[][]{Kjeldsen08,Schou18}.
\end{enumerate}
Figure~\ref{fig:prewhitening} shows the prewhitened power spectrum after extracting all 19 identified modes, where it can be seen that they account for most of the power within the p-mode frequency range.

The problem of mode identification, in the sense of assigning a pair $(n,\ell)$ to the extracted frequencies, may not be a trivial one to solve. There are numerous instances of seismic studies that led to uncertainty on the mode identification \citep[e.g.][]{CarrierEggenberger06,Appourchaux08,Bedding10}, which is made worse in the case of single-site ground-based observations. As an additional check on the above observational procedure, we adopted a model-based approach to mode identification based upon the work of \citet{White11_1,White11_2}, in order to verify whether the observed position of the $\ell\!=\!0$ ridge in the \'echelle diagram is consistent with expectations from stellar models. Based on the stellar model grid of \citet{Li23}, we calculated, for each model in that grid, a likelihood function, $\mathcal{L}_i\!\sim\!\exp(-\chi_i^2/2)$, where the discrepancy function, $\chi_i^2$, is given by the sum of the error-normalised discrepancies for $T_{\rm eff}$, [Fe/H], and $\Delta\nu$, adopted as observational constraints. We next computed a model prediction of the quantity $\varepsilon\,\Delta\nu$, which gives the absolute position of the $\ell\!=\!0$ ridge in an \'echelle diagram (cf.~Eq.~\ref{eq:asymptotic}). This was simply done by constructing a likelihood-weighted histogram of $\varepsilon\,\Delta\nu$ for the models in the grid (see bottom panel of Fig.~\ref{fig:echelle}). According to this, the power ridge just below $100\:{\rm \mu Hz}$ in the top panel of Fig.~\ref{fig:echelle} should correspond to $\ell\!=\!2,\!0$ (rather than $\ell\!=\!1$), thus providing support to our adopted mode identification. We note the presence of a small offset between the model-based and observed (shown as a vertical dotted line) $\varepsilon\,\Delta\nu$. This is to be expected, and is due to the fact that model frequencies were not corrected for the surface effect \citep[see e.g.][]{BallGizon14}. We will be investigating the magnitude of the surface effect in this $T_{\rm eff}$ regime in a follow-up study.

\subsection{Oscillation amplitudes}\label{sec:Amp}

The measured amplitudes of individual modes are affected by the stochastic nature of the excitation and damping. We hence followed the procedure described in \citet{Kjeldsen05,Kjeldsen08} to measure the oscillation amplitude envelope in a way that is independent of these effects. In short, we heavily smoothed the power spectrum by convolving it with a Gaussian having a full width at half maximum (FWHM) of $4\,\Delta\nu$; converted to power density; fitted and subtracted the background noise; and multiplied by $\Delta\nu/c$ and took the square root, thus converting to amplitude per radial mode \citep[a value of $c\!=\!4.09$ was adopted, representing the effective number of modes per order for full-disk velocity observations, normalised to the amplitudes of radial modes;][]{Kjeldsen08}.

The envelope peak amplitude thus obtained is $\varv_{\rm osc}\!=\!2.6\pm0.5\:{\rm cm\,s^{-1}}$, or a mere ${\sim} 14\%$ of the solar value \citep[$\varv_{{\rm osc},\sun}\!=\!18.7\:{\rm cm\,s^{-1}}$, as measured using stellar techniques and averaged over one full solar cycle;][]{Kjeldsen08}. The associated uncertainty was estimated as the standard deviation resulting from having applied the above procedure to the power spectra of 2000 artificial time series, generated using the asteroFLAG Artificial DataSet Generator, version 3 \citep[\texttt{AADG3};][]{Ball18}. Each simulated time series contained as input all extracted mode frequencies (cf.~Table~\ref{tab:freqs}), with mode lifetimes allowed to vary across simulations, and was sampled adopting the $\epsilon$~Indi observing window. A fixed $\nu_{\rm max}$ and amplitude per radial mode were assumed based on the corresponding measured quantities. The reported error bar thus takes into account different sources of uncertainty, namely realisation noise, the stochastic nature of the oscillations, and the white-noise level (whose input value was subject to a $10\%$ perturbation). Owing to the single-epoch nature of the observations, no attempt was made to quantify the uncertainty related to potential variations induced by the stellar activity cycle (see Fig.~\ref{fig:cycle}).

\begin{figure}[!t]
\centering
\includegraphics[width=0.5\textwidth,trim=0cm 0cm 0cm 0.2cm,clip]{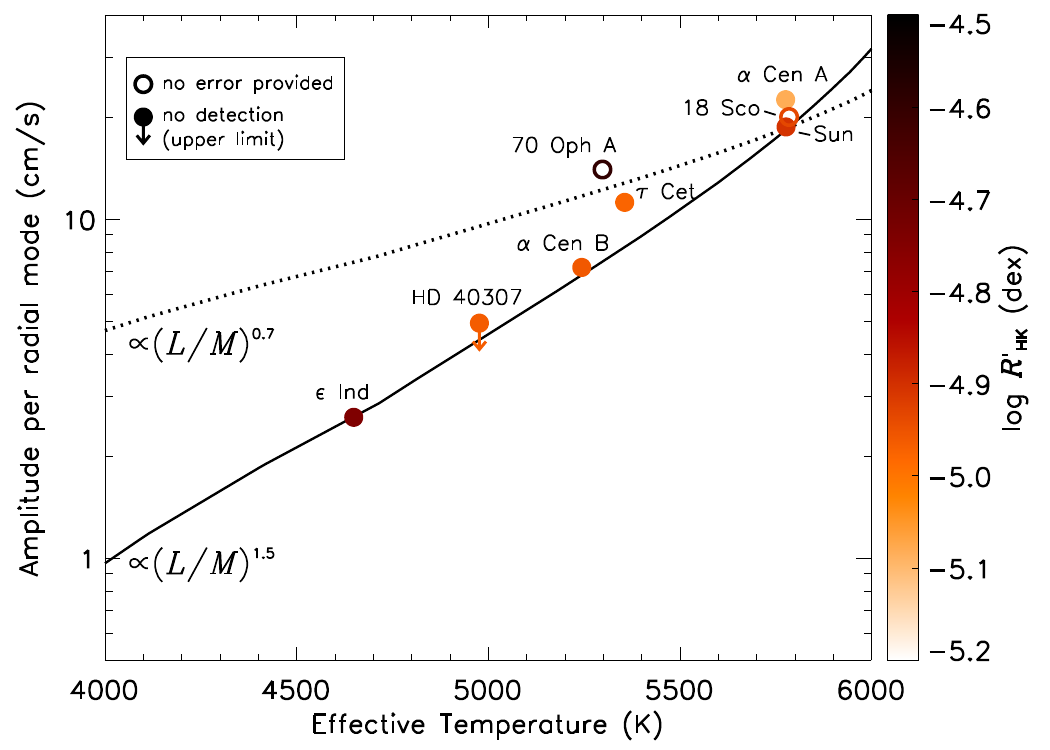}
\caption{Amplitude per radial mode as a function of $T_{\rm eff}$ for $\epsilon$~Indi and a number of cool dwarfs with published measurements. Reported statistical uncertainties for the mode amplitudes are comparable in size to the plotted symbols. Data points are colour-coded according to the corresponding $\log R^{\prime}_{\rm HK}$ ratio. Except for $\epsilon$~Indi (see Appendix~\ref{sec:Activity}) and the Sun \citep{MamajekHillenbrand08}, all $\log R^{\prime}_{\rm HK}$ measurements are from \citet{GomesdaSilva21}. Two scalings of the mode amplitudes are shown, differing in terms of the exponent $s$ ($s\!=\!0.7$, dotted curve; $s\!=\!1.5$, solid curve). The adopted $L/M$ relation is from stellar models in a 4.57 Gyr, solar-metallicity isochrone computed with the PAdova and TRieste Stellar Evolution Code \citep[\texttt{PARSEC};][]{PARSEC}.}
\label{fig:amp}
\end{figure}

Figure~\ref{fig:amp} shows the amplitude per radial mode as a function of $T_{\rm eff}$ (colour-coded according to the chromospheric emission ratio, $\log R^{\prime}_{\rm HK}$) for $\epsilon$~Indi and a number of cool dwarfs with published measurements. We note that the amount of smoothing of the power spectrum affects the exact height of the smoothed amplitude envelope, and hence the estimate of $\varv_{\rm osc}$. Measurements for the Sun and $\alpha$~Centauri~A and B \citep{Kjeldsen08}, as well as for $\tau$~Ceti \citep{Teixeira09} were obtained following the same procedure as described above. The value plotted for 70~Ophiuchi~A \citep{CarrierEggenberger06} corresponds to the upper bound on the amplitudes of the highest mode peaks detected and has no associated uncertainty, while an estimate (no error provided) of the mode amplitudes for the solar twin 18 Scorpii is given in \citet{Bazot11}. Finally, our reanalysis of the ESPRESSO GTO radial-velocity data of HD~40307 showed no p-mode detection (see Appendix~\ref{sec:HD40307}), and so the plotted value corresponds to an upper limit.

Based on calculations by \citet[][]{JCDFrandsen83}, \citet[][]{KjeldsenBedding95} suggested a scaling of the oscillation amplitudes of p modes in Doppler velocity of the form $\varv_{\rm osc}\!\propto\!(L/M)^s$, with $s\!=\!1$, and where $L$ and $M$ are respectively the stellar luminosity and mass. The numerical value of the exponent $s$ has since been revised theoretically, based on models of main-sequence stars, and found to lie in the range 0.7--1.5 \citep[see e.g.][]{Houdek99,Samadi05,Samadi07}. On the other hand, observational\footnote{Observational results draw mostly from studies conducted in photometry. Photometric mode amplitudes, once corrected to bolometric amplitudes, are expected to scale as $A_{\rm bol}\!\propto\!\varv_{\rm osc}\,T_{\rm eff}^{1-r}$ \citep[with $r\!=\!1.5$ if assuming adiabatic oscillations;][]{KjeldsenBedding95}, thus allowing for conversion between photometric and Doppler observations.} studies based on large ensembles of main-sequence and subgiant Kepler stars have constrained $s$ to the approximate range\footnote{The procedure described in \citet{Kjeldsen05,Kjeldsen08} has been widely implemented in automated analysis pipelines with the goal of calibrating scaling relations. \citet{Corsaro13} make exclusive use of amplitudes derived following this procedure. \citet{Verner11} compare different analysis pipelines, several of which adopt this procedure; for the sake of homogeneity, we consider only the latter pipelines
here  \citep[i.e.][]{Huber09,Hekker10,Mathur10}.} 0.5--1.0, depending on $T_{\rm eff}$ \citep[][]{Verner11,Corsaro13}.

We show, in Fig.~\ref{fig:amp}, two scalings of the mode amplitudes corresponding to the extrema of the theoretical range in $s$ (i.e. $s\!=\!0.7$ and 1.5). The displayed mode amplitude measurements hint at a transition to a scaling closer to $(L/M)^{1.5}$ below ${\sim} 5500\:{\rm K}$ \citep[cf.][where a negative ${\rm d}s/{\rm d}T_{\rm eff}$ gradient was determined]{Verner11}. This is supported by the mode amplitudes measured herein for $\epsilon$~Indi and the upper limit on the mode amplitudes obtained for HD~40307. The fraction of magnetically active stars among K dwarfs is higher than among G dwarfs \citep[e.g.][]{Jenkins11,GomesdaSilva21}. At the same time, increasing levels of activity are known to suppress the amplitudes of solar-like oscillations \citep[][]{Garcia10,Chaplin11,Bonanno14,Campante14}, which could therefore be the underlying cause for the apparent transition between scaling relations when moving down in $T_{\rm eff}$ in Fig.~\ref{fig:amp}. We find no significant correlation between $\log R^{\prime}_{\rm HK}$ and $T_{\rm eff}$ (Spearman's rank correlation coefficient, $\rho\!=\!-0.17$, and large $p$-value, $p\!\gg\!0.05$) for the displayed cool-dwarf sample. However, given the limited size of this sample, we refrain from making more general considerations regarding the role of activity in this context, and advocate for the inclusion of its effect in the calibration of the mode-amplitude scaling in this $T_{\rm eff}$ regime as more targets are observed. Finally, it is worth noting that $\epsilon$~Indi ($\log R^{\prime}_{\rm HK}\!=\!-4.742\pm0.004\:{\rm dex}$; see Appendix~\ref{sec:Activity}) and 70~Ophiuchi~A \citep[$\log R^{\prime}_{\rm HK}\!=\!-4.594\pm0.005\:{\rm dex}$;][]{GomesdaSilva21} are the only relatively active stars in a sample otherwise biased toward inactive stars ($-5.1\!<\!\log R^{\prime}_{\rm HK}\!<\!-4.9\:{\rm dex}$).

\subsection{Oscillation lifetimes}\label{sec:Lifetimes}

Solar-like oscillations are stochastically excited and damped by near-surface convection. The power spectrum of a single mode that is observed for long enough will appear as an erratic function concealing a Lorentzian profile, the width of which indicates the mode
lifetime \citep[e.g.][]{Anderson90}. If, as in the present case, the
observations are not long enough to fully resolve the Lorentzian
profile, then the effect of the finite mode lifetime is to randomly shift each mode peak from its true position by a
small amount \citep[e.g.][]{Bedding04,Kjeldsen05}. Measuring this scatter provides an opportunity to infer the mode lifetimes in $\epsilon$~Indi.

We inferred the mode lifetimes by measuring the scatter of the observed frequencies of radial modes (which are not impacted by rotation) about their power ridge in the \'echelle diagram and comparing with simulations (see Appendix~\ref{sec:Lifetimes_calibration}). The top panel of Fig.~\ref{fig:lifetime} shows the outcome of this calibration procedure for the $\epsilon$~Indi observing window. Although an upper bound on the mode lifetimes is weakly constrained, it is safe to say that lifetimes are at least a few days long ($\gtrsim\!3\:{\rm d}$). For context, the average mode lifetime in the Sun, measured in the range 2.8--$3.4\:{\rm mHz}$, is $2.88\pm0.07\:{\rm d}$ \citep[][]{Chaplin97}, being slightly longer than for $\tau$~Ceti \citep[$1.7\pm0.5\:{\rm d}$;][]{Teixeira09}, and in line with that measured for $\alpha$~Centauri~B \citep[$3.3^{+1.8}_{-0.9}\:{\rm d}$ at $3.6\:{\rm mHz}$ and $1.9^{+0.7}_{-0.4}\:{\rm d}$ at $4.6\:{\rm mHz}$;][]{Kjeldsen05}.

\section{Conclusion and outlook}\label{sec:Conclusions}

\begin{figure}[!t]
\centering
\includegraphics[width=0.5\textwidth,trim=0.8cm 0.8cm 0.1cm 1.5cm,clip]{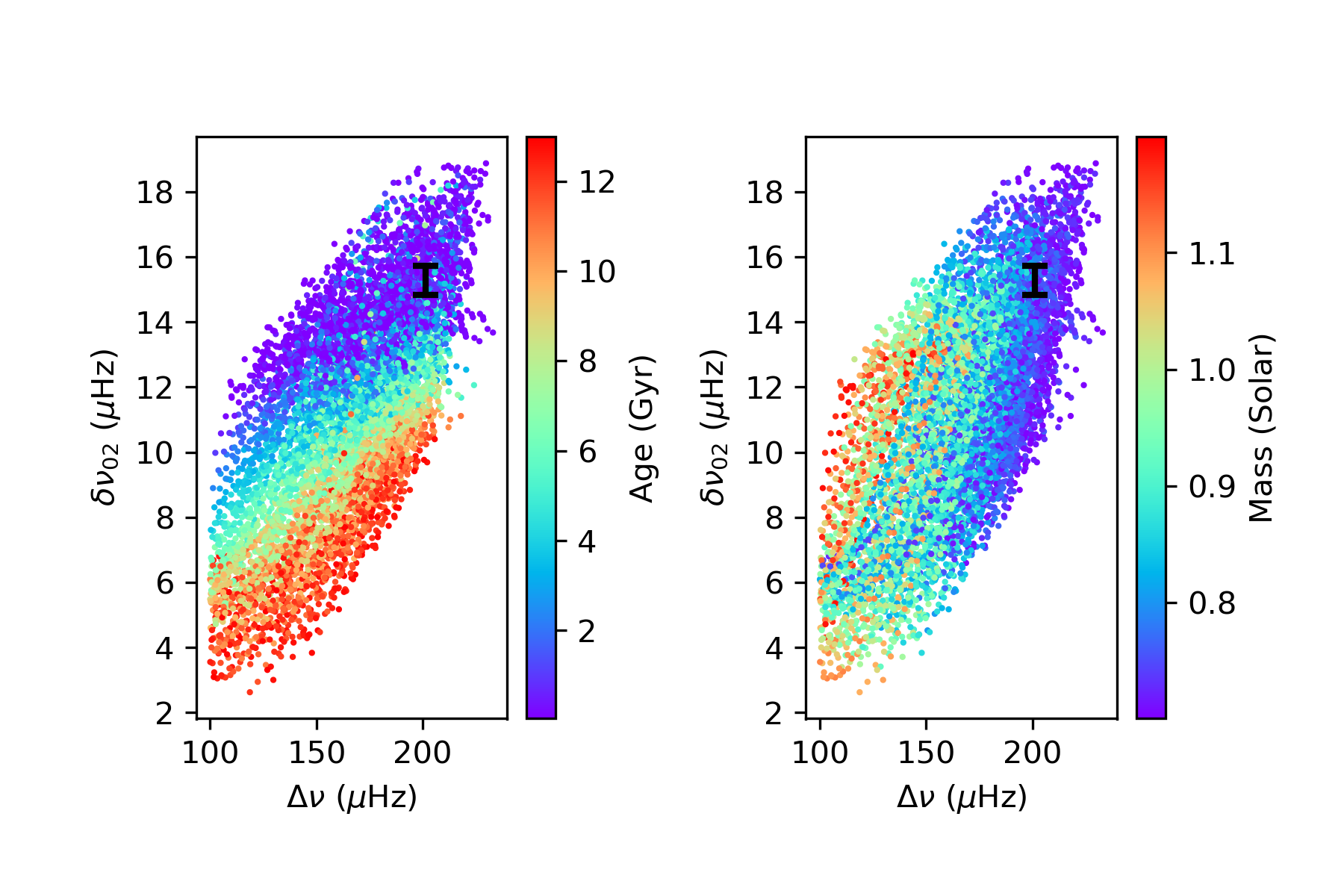}
\caption{C-D diagram, showing $\delta\nu_{02}$ vs~$\Delta\nu$. Models from the grid of \citet{Li23} are colour-coded according to age (left) and mass (right). The location of $\epsilon$~Indi is indicated by the black symbol in both panels. The error on $\Delta\nu$ is too small to be discerned.}
\label{fig:cd_diagram}
\end{figure}

In this letter we have presented the definite detection of solar-like oscillations in the bright K5 dwarf $\epsilon$~Indi based on radial-velocity observations carried out with the ESPRESSO spectrograph. This campaign hence unambiguously demonstrates the potential of ESPRESSO for cool-dwarf asteroseismology, effectively opening up a new observational domain in the field.

Measured mode amplitudes for $\epsilon$~Indi are approximately two times lower than predicted from a nominal $L/M$ scaling relation, favouring a scaling closer to $(L/M)^{1.5}$ below ${\sim} 5500\:{\rm K}$. A calibration of the mode-amplitude scaling relation in this $T_{\rm eff}$ regime is thus called for as more targets are observed for asteroseismology. Mode amplitudes are determined by a delicate balance between the energy supply and the mode damping, both being directly connected to the turbulent velocity field associated with convection \citep[][]{HoudekDupret15}. The measurement of oscillation modes in K dwarfs will hence allow us to constrain the dynamical coupling between pulsations and near-surface convection in a regime yet unexplored. Moreover, measured mode amplitudes, used in combination either with 1D non-local time-dependent convection models \citep[][]{Chaplin05} or with state-of-the-art 3D stellar atmosphere simulations \citep[][]{Zhou21}, will enable predictions of amplitudes in photometry. This information is key to accurately estimating the PLATO seismic yield \citep[][]{Miglio17,Goupil24} and can potentially influence the PLATO pipeline development strategy \citep[][]{Cunha21}.

Furthermore, $\epsilon$~Indi is the only known system containing T-type brown dwarfs for which a test of substellar cooling with time and a coevality test of model isochrones are both made possible \citep[][]{Chen22}. These tests would greatly benefit from having a precise seismic age for the host star, as it still remains a major source of uncertainty in the evolutionary and atmospheric modelling of the system. We display, in Fig.~\ref{fig:cd_diagram}, the location of $\epsilon$~Indi in a C-D diagram \citep[][]{JCD84}, showing $\delta\nu_{02}$ versus $\Delta\nu$. Inspection of this diagram implies a seismic stellar age $<\!4\:{\rm Gyr}$, consistent with most literature measurements \citep[see][and references therein]{Chen22}, which include activity-based estimates, as well as ages from kinematics and isochrone fitting. Detailed asteroseismic modelling of $\epsilon$~Indi will be the subject of a follow-up study.

\begin{acknowledgements}
We thank the anonymous referee for their valuable comments and attention to detail. Based on observations collected at the European Southern Observatory under ESO programmes 109.236P.001 ($\epsilon$~Indi; PI: Campante) and 0102.D-0346(A) (HD~40307; PI: Bouchy). We thank Fran\c{c}ois Bouchy for having shared with us the ESPRESSO GTO data of HD~40307 when these data were not yet public. This work was supported by Funda\c c\~ao para a Ci\^encia e a Tecnologia (FCT) through research grants UIDB/04434/2020 and UIDP/04434/2020. Co-funded by the European Union (ERC, FIERCE, 101052347). Views and opinions expressed are however those of the author(s) only and do not necessarily reflect those of the European Union or the European Research Council. Neither the European Union nor the granting authority can be held responsible for them. Funding for the Stellar Astrophysics Centre (SAC) was provided by the Danish National Research Foundation (grant agreement no.: DNRF106). TLC is supported by FCT in the form of a work contract (CEECIND/00476/2018). AMS acknowledges support from FCT through fellowship 2020.05387.BD. TRB is supported by the Australian Research Council (FL220100117). DLB gratefully acknowledges support from NASA (NNX16AB76G, 80NSSC22K0622) and the Whitaker Endowed Fund at Florida Gulf Coast University. MSC acknowledges support from FCT/MCTES through grants 2022.03993.PTDC and CEECIND/02619/2017. RAG acknowledges support from the GOLF and PLATO grants from Centre National d'{\'{E}}tudes Spatiales (CNES). MSL acknowledges support from VILLUM FONDEN (research grant 42101) and The Independent Research Fund Denmark's Inge Lehmann programme (grant agreement no.: 1131-00014B).
\end{acknowledgements}

\bibliographystyle{aa} 
\bibliography{main.bib} 

\begin{thebibliography}{84}
\expandafter\ifx\csname natexlab\endcsname\relax\def\natexlab#1{#1}\fi

\bibitem[{{Adibekyan} {et~al.}(2012){Adibekyan}, {Sousa}, {Santos}, {Delgado Mena}, {Gonz{\'a}lez Hern{\'a}ndez}, {Israelian}, {Mayor}, \& {Khachatryan}}]{Adibekyan12}
{Adibekyan}, V.~Z., {Sousa}, S.~G., {Santos}, N.~C., {et~al.} 2012, \aap, 545, A32

\bibitem[{{Aerts}(2021)}]{Aerts21}
{Aerts}, C. 2021, Reviews of Modern Physics, 93, 015001

\bibitem[{{Anderson} {et~al.}(1990){Anderson}, {Duvall}, \& {Jefferies}}]{Anderson90}
{Anderson}, E.~R., {Duvall}, Thomas~L., J., \& {Jefferies}, S.~M. 1990, \apj, 364, 699

\bibitem[{{Appourchaux} {et~al.}(2008){Appourchaux}, {Michel}, {Auvergne}, {Baglin}, {Toutain}, {Baudin}, {Benomar}, {Chaplin}, {Deheuvels}, {Samadi}, {Verner}, {Boumier}, {Garc{\'\i}a}, {Mosser}, {Hulot}, {Ballot}, {Barban}, {Elsworth}, {Jim{\'e}nez-Reyes}, {Kjeldsen}, {R{\'e}gulo}, \& {Roxburgh}}]{Appourchaux08}
{Appourchaux}, T., {Michel}, E., {Auvergne}, M., {et~al.} 2008, \aap, 488, 705

\bibitem[{{Arentoft} {et~al.}(2008){Arentoft}, {Kjeldsen}, {Bedding}, {Bazot}, {Christensen-Dalsgaard}, {Dall}, {Karoff}, {Carrier}, {Eggenberger}, {Sosnowska}, {Wittenmyer}, {Endl}, {Metcalfe}, {Hekker}, {Reffert}, {Butler}, {Bruntt}, {Kiss}, {O'Toole}, {Kambe}, {Ando}, {Izumiura}, {Sato}, {Hartmann}, {Hatzes}, {Bouchy}, {Mosser}, {Appourchaux}, {Barban}, {Berthomieu}, {Garcia}, {Michel}, {Provost}, {Turck-Chi{\`e}ze}, {Marti{\'c}}, {Lebrun}, {Schmitt}, {Bertaux}, {Bonanno}, {Benatti}, {Claudi}, {Cosentino}, {Leccia}, {Frandsen}, {Brogaard}, {Glowienka}, {Grundahl}, \& {Stempels}}]{Arentoft08}
{Arentoft}, T., {Kjeldsen}, H., {Bedding}, T.~R., {et~al.} 2008, \apj, 687, 1180

\bibitem[{{Baglin} {et~al.}(2006){Baglin}, {Auvergne}, {Barge}, {Deleuil}, {Catala}, {Michel}, {Weiss}, \& {COROT Team}}]{CoRoT}
{Baglin}, A., {Auvergne}, M., {Barge}, P., {et~al.} 2006, in ESA Special Publication, Vol. 1306, The CoRoT Mission Pre-Launch Status - Stellar Seismology and Planet Finding, ed. M.~{Fridlund}, A.~{Baglin}, J.~{Lochard}, \& L.~{Conroy}, 33

\bibitem[{{Ball} {et~al.}(2018){Ball}, {Chaplin}, {Schofield}, {Miglio}, {Bossini}, {Davies}, \& {Girardi}}]{Ball18}
{Ball}, W.~H., {Chaplin}, W.~J., {Schofield}, M., {et~al.} 2018, \apjs, 239, 34

\bibitem[{{Ball} \& {Gizon}(2014)}]{BallGizon14}
{Ball}, W.~H. \& {Gizon}, L. 2014, \aap, 568, A123

\bibitem[{{Bazot} {et~al.}(2011){Bazot}, {Ireland}, {Huber}, {Bedding}, {Broomhall}, {Campante}, {Carfantan}, {Chaplin}, {Elsworth}, {Mel{\'e}ndez}, {Petit}, {Th{\'e}ado}, {Van Grootel}, {Arentoft}, {Asplund}, {Castro}, {Christensen-Dalsgaard}, {Do Nascimento}, {Dintrans}, {Dumusque}, {Kjeldsen}, {McAlister}, {Metcalfe}, {Monteiro}, {Santos}, {Sousa}, {Sturmann}, {Sturmann}, {ten Brummelaar}, {Turner}, \& {Vauclair}}]{Bazot11}
{Bazot}, M., {Ireland}, M.~J., {Huber}, D., {et~al.} 2011, \aap, 526, L4

\bibitem[{{Bedding} {et~al.}(2004){Bedding}, {Kjeldsen}, {Butler}, {McCarthy}, {Marcy}, {O'Toole}, {Tinney}, \& {Wright}}]{Bedding04}
{Bedding}, T.~R., {Kjeldsen}, H., {Butler}, R.~P., {et~al.} 2004, \apj, 614, 380

\bibitem[{{Bedding} {et~al.}(2010){Bedding}, {Kjeldsen}, {Campante}, {Appourchaux}, {Bonanno}, {Chaplin}, {Garcia}, {Marti{\'c}}, {Mosser}, {Butler}, {Bruntt}, {Kiss}, {O'Toole}, {Kambe}, {Ando}, {Izumiura}, {Sato}, {Hartmann}, {Hatzes}, {Barban}, {Berthomieu}, {Michel}, {Provost}, {Turck-Chi{\`e}ze}, {Lebrun}, {Schmitt}, {Bertaux}, {Benatti}, {Claudi}, {Cosentino}, {Leccia}, {Frandsen}, {Brogaard}, {Glowienka}, {Grundahl}, {Stempels}, {Arentoft}, {Bazot}, {Christensen-Dalsgaard}, {Dall}, {Karoff}, {Lundgreen-Nielsen}, {Carrier}, {Eggenberger}, {Sosnowska}, {Wittenmyer}, {Endl}, {Metcalfe}, {Hekker}, \& {Reffert}}]{Bedding10}
{Bedding}, T.~R., {Kjeldsen}, H., {Campante}, T.~L., {et~al.} 2010, \apj, 713, 935

\bibitem[{{Bonanno} \& {Corsaro}(2022)}]{Bonanno22}
{Bonanno}, A. \& {Corsaro}, E. 2022, \apjl, 939, L26

\bibitem[{{Bonanno} {et~al.}(2014){Bonanno}, {Corsaro}, \& {Karoff}}]{Bonanno14}
{Bonanno}, A., {Corsaro}, E., \& {Karoff}, C. 2014, \aap, 571, A35

\bibitem[{{Borucki} {et~al.}(2010){Borucki}, {Koch}, {Basri}, {Batalha}, {Brown}, {Caldwell}, {Caldwell}, {Christensen-Dalsgaard}, {Cochran}, {DeVore}, {Dunham}, {Dupree}, {Gautier}, {Geary}, {Gilliland}, {Gould}, {Howell}, {Jenkins}, {Kondo}, {Latham}, {Marcy}, {Meibom}, {Kjeldsen}, {Lissauer}, {Monet}, {Morrison}, {Sasselov}, {Tarter}, {Boss}, {Brownlee}, {Owen}, {Buzasi}, {Charbonneau}, {Doyle}, {Fortney}, {Ford}, {Holman}, {Seager}, {Steffen}, {Welsh}, {Rowe}, {Anderson}, {Buchhave}, {Ciardi}, {Walkowicz}, {Sherry}, {Horch}, {Isaacson}, {Everett}, {Fischer}, {Torres}, {Johnson}, {Endl}, {MacQueen}, {Bryson}, {Dotson}, {Haas}, {Kolodziejczak}, {Van Cleve}, {Chandrasekaran}, {Twicken}, {Quintana}, {Clarke}, {Allen}, {Li}, {Wu}, {Tenenbaum}, {Verner}, {Bruhweiler}, {Barnes}, \& {Prsa}}]{Kepler}
{Borucki}, W.~J., {Koch}, D., {Basri}, G., {et~al.} 2010, Science, 327, 977

\bibitem[{{Bressan} {et~al.}(2012){Bressan}, {Marigo}, {Girardi}, {Salasnich}, {Dal Cero}, {Rubele}, \& {Nanni}}]{PARSEC}
{Bressan}, A., {Marigo}, P., {Girardi}, L., {et~al.} 2012, \mnras, 427, 127

\bibitem[{{Brown} {et~al.}(1991){Brown}, {Gilliland}, {Noyes}, \& {Ramsey}}]{Brown91}
{Brown}, T.~M., {Gilliland}, R.~L., {Noyes}, R.~W., \& {Ramsey}, L.~W. 1991, \apj, 368, 599

\bibitem[{{Campante} {et~al.}(2015){Campante}, {Barclay}, {Swift}, {Huber}, {Adibekyan}, {Cochran}, {Burke}, {Isaacson}, {Quintana}, {Davies}, {Silva Aguirre}, {Ragozzine}, {Riddle}, {Baranec}, {Basu}, {Chaplin}, {Christensen-Dalsgaard}, {Metcalfe}, {Bedding}, {Handberg}, {Stello}, {Brewer}, {Hekker}, {Karoff}, {Kolbl}, {Law}, {Lundkvist}, {Miglio}, {Rowe}, {Santos}, {Van Laerhoven}, {Arentoft}, {Elsworth}, {Fischer}, {Kawaler}, {Kjeldsen}, {Lund}, {Marcy}, {Sousa}, {Sozzetti}, \& {White}}]{Kepler444}
{Campante}, T.~L., {Barclay}, T., {Swift}, J.~J., {et~al.} 2015, \apj, 799, 170

\bibitem[{{Campante} {et~al.}(2014){Campante}, {Chaplin}, {Lund}, {Huber}, {Hekker}, {Garc{\'\i}a}, {Corsaro}, {Handberg}, {Miglio}, {Arentoft}, {Basu}, {Bedding}, {Christensen-Dalsgaard}, {Davies}, {Elsworth}, {Gilliland}, {Karoff}, {Kawaler}, {Kjeldsen}, {Lundkvist}, {Metcalfe}, {Silva Aguirre}, \& {Stello}}]{Campante14}
{Campante}, T.~L., {Chaplin}, W.~J., {Lund}, M.~N., {et~al.} 2014, \apj, 783, 123

\bibitem[{{Carrier} \& {Eggenberger}(2006)}]{CarrierEggenberger06}
{Carrier}, F. \& {Eggenberger}, P. 2006, \aap, 450, 695

\bibitem[{{Chaplin} {et~al.}(2011){Chaplin}, {Bedding}, {Bonanno}, {Broomhall}, {Garc{\'\i}a}, {Hekker}, {Huber}, {Verner}, {Basu}, {Elsworth}, {Houdek}, {Mathur}, {Mosser}, {New}, {Stevens}, {Appourchaux}, {Karoff}, {Metcalfe}, {Molenda-{\.Z}akowicz}, {Monteiro}, {Thompson}, {Christensen-Dalsgaard}, {Gilliland}, {Kawaler}, {Kjeldsen}, {Ballot}, {Benomar}, {Corsaro}, {Campante}, {Gaulme}, {Hale}, {Handberg}, {Jarvis}, {R{\'e}gulo}, {Roxburgh}, {Salabert}, {Stello}, {Mullally}, {Li}, \& {Wohler}}]{Chaplin11}
{Chaplin}, W.~J., {Bedding}, T.~R., {Bonanno}, A., {et~al.} 2011, \apjl, 732, L5

\bibitem[{{Chaplin} {et~al.}(1997){Chaplin}, {Elsworth}, {Isaak}, {McLeod}, {Miller}, \& {New}}]{Chaplin97}
{Chaplin}, W.~J., {Elsworth}, Y., {Isaak}, G.~R., {et~al.} 1997, \mnras, 288, 623

\bibitem[{{Chaplin} {et~al.}(2005){Chaplin}, {Houdek}, {Elsworth}, {Gough}, {Isaak}, \& {New}}]{Chaplin05}
{Chaplin}, W.~J., {Houdek}, G., {Elsworth}, Y., {et~al.} 2005, \mnras, 360, 859

\bibitem[{{Chen} {et~al.}(2022){Chen}, {Li}, {Brandt}, {Dupuy}, {Cardoso}, \& {McCaughrean}}]{Chen22}
{Chen}, M., {Li}, Y., {Brandt}, T.~D., {et~al.} 2022, \aj, 163, 288

\bibitem[{{Christensen-Dalsgaard}(1984)}]{JCD84}
{Christensen-Dalsgaard}, J. 1984, in Space Research in Stellar Activity and Variability, ed. A.~{Mangeney} \& F.~{Praderie}, 11

\bibitem[{{Christensen-Dalsgaard} \& {Frandsen}(1983)}]{JCDFrandsen83}
{Christensen-Dalsgaard}, J. \& {Frandsen}, S. 1983, \solphys, 82, 469

\bibitem[{{Corsaro} \& {De Ridder}(2014)}]{DIAMONDS}
{Corsaro}, E. \& {De Ridder}, J. 2014, \aap, 571, A71

\bibitem[{{Corsaro} {et~al.}(2013){Corsaro}, {Fr{\"o}hlich}, {Bonanno}, {Huber}, {Bedding}, {Benomar}, {De Ridder}, \& {Stello}}]{Corsaro13}
{Corsaro}, E., {Fr{\"o}hlich}, H.~E., {Bonanno}, A., {et~al.} 2013, \mnras, 430, 2313

\bibitem[{{Cunha} {et~al.}(2021){Cunha}, {Roxburgh}, {Aguirre B{\o}rsen-Koch}, {Ball}, {Basu}, {Chaplin}, {Goupil}, {Nsamba}, {Ong}, {Reese}, {Verma}, {Belkacem}, {Campante}, {Christensen-Dalsgaard}, {Clara}, {Deheuvels}, {Monteiro}, {Noll}, {Ouazzani}, {R{\o}rsted}, {Stokholm}, \& {Winther}}]{Cunha21}
{Cunha}, M.~S., {Roxburgh}, I.~W., {Aguirre B{\o}rsen-Koch}, V., {et~al.} 2021, \mnras, 508, 5864

\bibitem[{{Davies} {et~al.}(2014){Davies}, {Handberg}, {Miglio}, {Campante}, {Chaplin}, \& {Elsworth}}]{Davies14}
{Davies}, G.~R., {Handberg}, R., {Miglio}, A., {et~al.} 2014, \mnras, 445, L94

\bibitem[{{Dekker} {et~al.}(2000){Dekker}, {D'Odorico}, {Kaufer}, {Delabre}, \& {Kotzlowski}}]{UVES}
{Dekker}, H., {D'Odorico}, S., {Kaufer}, A., {Delabre}, B., \& {Kotzlowski}, H. 2000, in Society of Photo-Optical Instrumentation Engineers (SPIE) Conference Series, Vol. 4008, Optical and IR Telescope Instrumentation and Detectors, ed. M.~{Iye} \& A.~F. {Moorwood}, 534--545

\bibitem[{{Delgado Mena} {et~al.}(2021){Delgado Mena}, {Adibekyan}, {Santos}, {Tsantaki}, {Gonz{\'a}lez Hern{\'a}ndez}, {Sousa}, \& {Bertr{\'a}n de Lis}}]{DelgadoMena21}
{Delgado Mena}, E., {Adibekyan}, V., {Santos}, N.~C., {et~al.} 2021, \aap, 655, A99

\bibitem[{{Feng} {et~al.}(2019){Feng}, {Anglada-Escud{\'e}}, {Tuomi}, {Jones}, {Chanam{\'e}}, {Butler}, \& {Janson}}]{Feng19}
{Feng}, F., {Anglada-Escud{\'e}}, G., {Tuomi}, M., {et~al.} 2019, \mnras, 490, 5002

\bibitem[{{Garc{\'\i}a} {et~al.}(2010){Garc{\'\i}a}, {Mathur}, {Salabert}, {Ballot}, {R{\'e}gulo}, {Metcalfe}, \& {Baglin}}]{Garcia10}
{Garc{\'\i}a}, R.~A., {Mathur}, S., {Salabert}, D., {et~al.} 2010, Science, 329, 1032

\bibitem[{{Gomes da Silva} {et~al.}(2018){Gomes da Silva}, {Figueira}, {Santos}, \& {Faria}}]{ACTIN}
{Gomes da Silva}, J., {Figueira}, P., {Santos}, N., \& {Faria}, J. 2018, The Journal of Open Source Software, 3, 667

\bibitem[{{Gomes da Silva} {et~al.}(2021){Gomes da Silva}, {Santos}, {Adibekyan}, {Sousa}, {Campante}, {Figueira}, {Bossini}, {Delgado-Mena}, {Monteiro}, {de Laverny}, {Recio-Blanco}, \& {Lovis}}]{GomesdaSilva21}
{Gomes da Silva}, J., {Santos}, N.~C., {Adibekyan}, V., {et~al.} 2021, \aap, 646, A77

\bibitem[{{Gould} {et~al.}(2015){Gould}, {Huber}, {Penny}, \& {Stello}}]{Gould15}
{Gould}, A., {Huber}, D., {Penny}, M., \& {Stello}, D. 2015, Journal of Korean Astronomical Society, 48, 93

\bibitem[{{Goupil} {et~al.}(2024){Goupil}, {Catala}, {Samadi}, {Belkacem}, {Ouazzani}, {Reese}, {Appourchaux}, {Mathur}, {Cabrera}, {B{\"o}rner}, {Paproth}, {Moedas}, {Verma}, {Lebreton}, {Deal}, {Ballot}, {Chaplin}, {Christensen-Dalsgaard}, {Cunha}, {Lanza}, {Miglio}, {Morel}, {Serenelli}, {Mosser}, {Creevey}, {Moya}, {Garcia}, {Nielsen}, \& {Hatt}}]{Goupil24}
{Goupil}, M.~J., {Catala}, C., {Samadi}, R., {et~al.} 2024, arXiv e-prints, arXiv:2401.07984

\bibitem[{{Grundahl} {et~al.}(2007){Grundahl}, {Kjeldsen}, {Christensen-Dalsgaard}, {Arentoft}, \& {Frandsen}}]{Grundahl07}
{Grundahl}, F., {Kjeldsen}, H., {Christensen-Dalsgaard}, J., {Arentoft}, T., \& {Frandsen}, S. 2007, Communications in Asteroseismology, 150, 300

\bibitem[{{Harvey}(1988)}]{Harvey88}
{Harvey}, J.~W. 1988, in Advances in Helio- and Asteroseismology, ed. J.~{Christensen-Dalsgaard} \& S.~{Frandsen}, Vol. 123, 497

\bibitem[{{Hatt} {et~al.}(2023){Hatt}, {Nielsen}, {Chaplin}, {Ball}, {Davies}, {Bedding}, {Buzasi}, {Chontos}, {Huber}, {Kayhan}, {Li}, {White}, {Cheng}, {Metcalfe}, \& {Stello}}]{Hatt23}
{Hatt}, E., {Nielsen}, M.~B., {Chaplin}, W.~J., {et~al.} 2023, \aap, 669, A67

\bibitem[{{Hekker} {et~al.}(2010){Hekker}, {Broomhall}, {Chaplin}, {Elsworth}, {Fletcher}, {New}, {Arentoft}, {Quirion}, \& {Kjeldsen}}]{Hekker10}
{Hekker}, S., {Broomhall}, A.~M., {Chaplin}, W.~J., {et~al.} 2010, \mnras, 402, 2049

\bibitem[{{Hon} {et~al.}(2021){Hon}, {Huber}, {Kuszlewicz}, {Stello}, {Sharma}, {Tayar}, {Zinn}, {Vrard}, \& {Pinsonneault}}]{Hon21}
{Hon}, M., {Huber}, D., {Kuszlewicz}, J.~S., {et~al.} 2021, \apj, 919, 131

\bibitem[{{Houdek} {et~al.}(1999){Houdek}, {Balmforth}, {Christensen-Dalsgaard}, \& {Gough}}]{Houdek99}
{Houdek}, G., {Balmforth}, N.~J., {Christensen-Dalsgaard}, J., \& {Gough}, D.~O. 1999, \aap, 351, 582

\bibitem[{{Houdek} \& {Dupret}(2015)}]{HoudekDupret15}
{Houdek}, G. \& {Dupret}, M.-A. 2015, Living Reviews in Solar Physics, 12, 8

\bibitem[{{Howell} {et~al.}(2014){Howell}, {Sobeck}, {Haas}, {Still}, {Barclay}, {Mullally}, {Troeltzsch}, {Aigrain}, {Bryson}, {Caldwell}, {Chaplin}, {Cochran}, {Huber}, {Marcy}, {Miglio}, {Najita}, {Smith}, {Twicken}, \& {Fortney}}]{K2}
{Howell}, S.~B., {Sobeck}, C., {Haas}, M., {et~al.} 2014, \pasp, 126, 398

\bibitem[{{Huber} {et~al.}(2009){Huber}, {Stello}, {Bedding}, {Chaplin}, {Arentoft}, {Quirion}, \& {Kjeldsen}}]{Huber09}
{Huber}, D., {Stello}, D., {Bedding}, T.~R., {et~al.} 2009, Communications in Asteroseismology, 160, 74

\bibitem[{Jeffreys(1961)}]{Jeffreys61}
Jeffreys, H. 1961, Theory of Probability, International series of monographs on physics (Clarendon Press)

\bibitem[{{Jenkins} {et~al.}(2011){Jenkins}, {Murgas}, {Rojo}, {Jones}, {Day-Jones}, {Jones}, {Clarke}, {Ruiz}, \& {Pinfield}}]{Jenkins11}
{Jenkins}, J.~S., {Murgas}, F., {Rojo}, P., {et~al.} 2011, \aap, 531, A8

\bibitem[{{Kjeldsen} \& {Bedding}(1995)}]{KjeldsenBedding95}
{Kjeldsen}, H. \& {Bedding}, T.~R. 1995, \aap, 293, 87

\bibitem[{{Kjeldsen} {et~al.}(2008){Kjeldsen}, {Bedding}, {Arentoft}, {Butler}, {Dall}, {Karoff}, {Kiss}, {Tinney}, \& {Chaplin}}]{Kjeldsen08}
{Kjeldsen}, H., {Bedding}, T.~R., {Arentoft}, T., {et~al.} 2008, \apj, 682, 1370

\bibitem[{{Kjeldsen} {et~al.}(2005){Kjeldsen}, {Bedding}, {Butler}, {Christensen-Dalsgaard}, {Kiss}, {McCarthy}, {Marcy}, {Tinney}, \& {Wright}}]{Kjeldsen05}
{Kjeldsen}, H., {Bedding}, T.~R., {Butler}, R.~P., {et~al.} 2005, \apj, 635, 1281

\bibitem[{{Kjeldsen} {et~al.}(1995){Kjeldsen}, {Bedding}, {Viskum}, \& {Frandsen}}]{Kjeldsen95}
{Kjeldsen}, H., {Bedding}, T.~R., {Viskum}, M., \& {Frandsen}, S. 1995, \aj, 109, 1313

\bibitem[{{Li} {et~al.}(2023){Li}, {Bedding}, {Stello}, {Huber}, {Hon}, {Joyce}, {Li}, {Perkins}, {White}, {Zinn}, {Howard}, {Isaacson}, {Hey}, \& {Kjeldsen}}]{Li23}
{Li}, Y., {Bedding}, T.~R., {Stello}, D., {et~al.} 2023, \mnras, 523, 916

\bibitem[{{Lillo-Box} {et~al.}(2022){Lillo-Box}, {Santos}, {Santerne}, {Silva}, {Barrado}, {Faria}, {Castro-Gonz{\'a}lez}, {Balsalobre-Ruza}, {Morales-Calder{\'o}n}, {Saavedra}, {Marfil}, {Sousa}, {Adibekyan}, {Berihuete}, {Barros}, {Delgado-Mena}, {Hu{\'e}lamo}, {Deleuil}, {Demangeon}, {Figueira}, {Grouffal}, {Aceituno}, {Azzaro}, {Bergond}, {Fern{\'a}ndez-Mart{\'\i}n}, {Galad{\'\i}}, {Gallego}, {Gardini}, {G{\'o}ngora}, {Guijarro}, {Hermelo}, {Mart{\'\i}n}, {M{\'\i}nguez}, {Montoya}, {Pedraz}, \& {Vico Linares}}]{KOBE}
{Lillo-Box}, J., {Santos}, N.~C., {Santerne}, A., {et~al.} 2022, \aap, 667, A102

\bibitem[{{Lund} {et~al.}(2017){Lund}, {Silva Aguirre}, {Davies}, {Chaplin}, {Christensen-Dalsgaard}, {Houdek}, {White}, {Bedding}, {Ball}, {Huber}, {Antia}, {Lebreton}, {Latham}, {Handberg}, {Verma}, {Basu}, {Casagrande}, {Justesen}, {Kjeldsen}, \& {Mosumgaard}}]{Lund17}
{Lund}, M.~N., {Silva Aguirre}, V., {Davies}, G.~R., {et~al.} 2017, \apj, 835, 172

\bibitem[{{Mamajek} \& {Stapelfeldt}(2023)}]{HWO}
{Mamajek}, E. \& {Stapelfeldt}, K. 2023, in American Astronomical Society Meeting Abstracts, Vol.~55, American Astronomical Society Meeting Abstracts, 116.07

\bibitem[{{Mamajek} \& {Hillenbrand}(2008)}]{MamajekHillenbrand08}
{Mamajek}, E.~E. \& {Hillenbrand}, L.~A. 2008, \apj, 687, 1264

\bibitem[{{Mathur} {et~al.}(2010){Mathur}, {Garc{\'\i}a}, {R{\'e}gulo}, {Creevey}, {Ballot}, {Salabert}, {Arentoft}, {Quirion}, {Chaplin}, \& {Kjeldsen}}]{Mathur10}
{Mathur}, S., {Garc{\'\i}a}, R.~A., {R{\'e}gulo}, C., {et~al.} 2010, \aap, 511, A46

\bibitem[{{Mathur} {et~al.}(2017){Mathur}, {Huber}, {Batalha}, {Ciardi}, {Bastien}, {Bieryla}, {Buchhave}, {Cochran}, {Endl}, {Esquerdo}, {Furlan}, {Howard}, {Howell}, {Isaacson}, {Latham}, {MacQueen}, \& {Silva}}]{Mathur17}
{Mathur}, S., {Huber}, D., {Batalha}, N.~M., {et~al.} 2017, \apjs, 229, 30

\bibitem[{{Mayor} {et~al.}(2003){Mayor}, {Pepe}, {Queloz}, {Bouchy}, {Rupprecht}, {Lo Curto}, {Avila}, {Benz}, {Bertaux}, {Bonfils}, {Dall}, {Dekker}, {Delabre}, {Eckert}, {Fleury}, {Gilliotte}, {Gojak}, {Guzman}, {Kohler}, {Lizon}, {Longinotti}, {Lovis}, {Megevand}, {Pasquini}, {Reyes}, {Sivan}, {Sosnowska}, {Soto}, {Udry}, {van Kesteren}, {Weber}, \& {Weilenmann}}]{HARPS}
{Mayor}, M., {Pepe}, F., {Queloz}, D., {et~al.} 2003, The Messenger, 114, 20

\bibitem[{{McCaughrean} {et~al.}(2004){McCaughrean}, {Close}, {Scholz}, {Lenzen}, {Biller}, {Brandner}, {Hartung}, \& {Lodieu}}]{McCaughrean04}
{McCaughrean}, M.~J., {Close}, L.~M., {Scholz}, R.~D., {et~al.} 2004, \aap, 413, 1029

\bibitem[{{Miglio} {et~al.}(2017){Miglio}, {Chiappini}, {Mosser}, {Davies}, {Freeman}, {Girardi}, {Jofr{\'e}}, {Kawata}, {Rendle}, {Valentini}, {Casagrande}, {Chaplin}, {Gilmore}, {Hawkins}, {Holl}, {Appourchaux}, {Belkacem}, {Bossini}, {Brogaard}, {Goupil}, {Montalb{\'a}n}, {Noels}, {Anders}, {Rodrigues}, {Piotto}, {Pollacco}, {Rauer}, {Allende Prieto}, {Avelino}, {Babusiaux}, {Barban}, {Barbuy}, {Basu}, {Baudin}, {Benomar}, {Bienaym{\'e}}, {Binney}, {Bland-Hawthorn}, {Bressan}, {Cacciari}, {Campante}, {Cassisi}, {Christensen-Dalsgaard}, {Combes}, {Creevey}, {Cunha}, {Jong}, {Laverny}, {Degl'Innocenti}, {Deheuvels}, {Depagne}, {Ridder}, {Matteo}, {Di Mauro}, {Dupret}, {Eggenberger}, {Elsworth}, {Famaey}, {Feltzing}, {Garc{\'\i}a}, {Gerhard}, {Gibson}, {Gizon}, {Haywood}, {Handberg}, {Heiter}, {Hekker}, {Huber}, {Ibata}, {Katz}, {Kawaler}, {Kjeldsen}, {Kurtz}, {Lagarde}, {Lebreton}, {Lund}, {Majewski}, {Marigo}, {Martig}, {Mathur}, {Minchev}, {Morel}, {Ortolani}, {Pinsonneault}, {Plez}, {Moroni}, {Pricopi},
  {Recio-Blanco}, {Reyl{\'e}}, {Robin}, {Roxburgh}, {Salaris}, {Santiago}, {Schiavon}, {Serenelli}, {Sharma}, {Aguirre}, {Soubiran}, {Steinmetz}, {Stello}, {Strassmeier}, {Ventura}, {Ventura}, {Walton}, \& {Worley}}]{Miglio17}
{Miglio}, A., {Chiappini}, C., {Mosser}, B., {et~al.} 2017, Astronomische Nachrichten, 338, 644

\bibitem[{{Pathak} {et~al.}(2021){Pathak}, {Petit dit de la Roche}, {Kasper}, {Sterzik}, {Absil}, {Boehle}, {Feng}, {Ivanov}, {Janson}, {Jones}, {Kaufer}, {K{\"a}ufl}, {Maire}, {Meyer}, {Pantin}, {Siebenmorgen}, {van den Ancker}, \& {Viswanath}}]{Pathak21}
{Pathak}, P., {Petit dit de la Roche}, D.~J.~M., {Kasper}, M., {et~al.} 2021, \aap, 652, A121

\bibitem[{{Pepe} {et~al.}(2021){Pepe}, {Cristiani}, {Rebolo}, {Santos}, {Dekker}, {Cabral}, {Di Marcantonio}, {Figueira}, {Lo Curto}, {Lovis}, {Mayor}, {M{\'e}gevand}, {Molaro}, {Riva}, {Zapatero Osorio}, {Amate}, {Manescau}, {Pasquini}, {Zerbi}, {Adibekyan}, {Abreu}, {Affolter}, {Alibert}, {Aliverti}, {Allart}, {Allende Prieto}, {{\'A}lvarez}, {Alves}, {Avila}, {Baldini}, {Bandy}, {Barros}, {Benz}, {Bianco}, {Borsa}, {Bourrier}, {Bouchy}, {Broeg}, {Calderone}, {Cirami}, {Coelho}, {Conconi}, {Coretti}, {Cumani}, {Cupani}, {D'Odorico}, {Damasso}, {Deiries}, {Delabre}, {Demangeon}, {Dumusque}, {Ehrenreich}, {Faria}, {Fragoso}, {Genolet}, {Genoni}, {G{\'e}nova Santos}, {Gonz{\'a}lez Hern{\'a}ndez}, {Hughes}, {Iwert}, {Kerber}, {Knudstrup}, {Landoni}, {Lavie}, {Lillo-Box}, {Lizon}, {Maire}, {Martins}, {Mehner}, {Micela}, {Modigliani}, {Monteiro}, {Monteiro}, {Moschetti}, {Murphy}, {Nunes}, {Oggioni}, {Oliveira}, {Oshagh}, {Pall{\'e}}, {Pariani}, {Poretti}, {Rasilla}, {Rebord{\~a}o}, {Redaelli}, {Santana Tschudi},
  {Santin}, {Santos}, {S{\'e}gransan}, {Schmidt}, {Segovia}, {Sosnowska}, {Sozzetti}, {Sousa}, {Span{\`o}}, {Su{\'a}rez Mascare{\~n}o}, {Tabernero}, {Tenegi}, {Udry}, \& {Zanutta}}]{ESPRESSO}
{Pepe}, F., {Cristiani}, S., {Rebolo}, R., {et~al.} 2021, \aap, 645, A96

\bibitem[{{Rains} {et~al.}(2020){Rains}, {Ireland}, {White}, {Casagrande}, \& {Karovicova}}]{Rains20}
{Rains}, A.~D., {Ireland}, M.~J., {White}, T.~R., {Casagrande}, L., \& {Karovicova}, I. 2020, \mnras, 493, 2377

\bibitem[{{Rauer} {et~al.}(2014){Rauer}, {Catala}, {Aerts}, {Appourchaux}, {Benz}, {Brandeker}, {Christensen-Dalsgaard}, {Deleuil}, {Gizon}, {Goupil}, {G{\"u}del}, {Janot-Pacheco}, {Mas-Hesse}, {Pagano}, {Piotto}, {Pollacco}, {Santos}, {Smith}, {Su{\'a}rez}, {Szab{\'o}}, {Udry}, {Adibekyan}, {Alibert}, {Almenara}, {Amaro-Seoane}, {Eiff}, {Asplund}, {Antonello}, {Barnes}, {Baudin}, {Belkacem}, {Bergemann}, {Bihain}, {Birch}, {Bonfils}, {Boisse}, {Bonomo}, {Borsa}, {Brand{\~a}o}, {Brocato}, {Brun}, {Burleigh}, {Burston}, {Cabrera}, {Cassisi}, {Chaplin}, {Charpinet}, {Chiappini}, {Church}, {Csizmadia}, {Cunha}, {Damasso}, {Davies}, {Deeg}, {D{\'\i}az}, {Dreizler}, {Dreyer}, {Eggenberger}, {Ehrenreich}, {Eigm{\"u}ller}, {Erikson}, {Farmer}, {Feltzing}, {de Oliveira Fialho}, {Figueira}, {Forveille}, {Fridlund}, {Garc{\'\i}a}, {Giommi}, {Giuffrida}, {Godolt}, {Gomes da Silva}, {Granzer}, {Grenfell}, {Grotsch-Noels}, {G{\"u}nther}, {Haswell}, {Hatzes}, {H{\'e}brard}, {Hekker}, {Helled}, {Heng}, {Jenkins},
  {Johansen}, {Khodachenko}, {Kislyakova}, {Kley}, {Kolb}, {Krivova}, {Kupka}, {Lammer}, {Lanza}, {Lebreton}, {Magrin}, {Marcos-Arenal}, {Marrese}, {Marques}, {Martins}, {Mathis}, {Mathur}, {Messina}, {Miglio}, {Montalban}, {Montalto}, {Monteiro}, {Moradi}, {Moravveji}, {Mordasini}, {Morel}, {Mortier}, {Nascimbeni}, {Nelson}, {Nielsen}, {Noack}, {Norton}, {Ofir}, {Oshagh}, {Ouazzani}, {P{\'a}pics}, {Parro}, {Petit}, {Plez}, {Poretti}, {Quirrenbach}, {Ragazzoni}, {Raimondo}, {Rainer}, {Reese}, {Redmer}, {Reffert}, {Rojas-Ayala}, {Roxburgh}, {Salmon}, {Santerne}, {Schneider}, {Schou}, {Schuh}, {Schunker}, {Silva-Valio}, {Silvotti}, {Skillen}, {Snellen}, {Sohl}, {Sousa}, {Sozzetti}, {Stello}, {Strassmeier}, {{\v{S}}vanda}, {Szab{\'o}}, {Tkachenko}, {Valencia}, {Van Grootel}, {Vauclair}, {Ventura}, {Wagner}, {Walton}, {Weingrill}, {Werner}, {Wheatley}, \& {Zwintz}}]{PLATO}
{Rauer}, H., {Catala}, C., {Aerts}, C., {et~al.} 2014, Experimental Astronomy, 38, 249

\bibitem[{{Ricker} {et~al.}(2015){Ricker}, {Winn}, {Vanderspek}, {Latham}, {Bakos}, {Bean}, {Berta-Thompson}, {Brown}, {Buchhave}, {Butler}, {Butler}, {Chaplin}, {Charbonneau}, {Christensen-Dalsgaard}, {Clampin}, {Deming}, {Doty}, {De Lee}, {Dressing}, {Dunham}, {Endl}, {Fressin}, {Ge}, {Henning}, {Holman}, {Howard}, {Ida}, {Jenkins}, {Jernigan}, {Johnson}, {Kaltenegger}, {Kawai}, {Kjeldsen}, {Laughlin}, {Levine}, {Lin}, {Lissauer}, {MacQueen}, {Marcy}, {McCullough}, {Morton}, {Narita}, {Paegert}, {Palle}, {Pepe}, {Pepper}, {Quirrenbach}, {Rinehart}, {Sasselov}, {Sato}, {Seager}, {Sozzetti}, {Stassun}, {Sullivan}, {Szentgyorgyi}, {Torres}, {Udry}, \& {Villasenor}}]{TESS}
{Ricker}, G.~R., {Winn}, J.~N., {Vanderspek}, R., {et~al.} 2015, Journal of Astronomical Telescopes, Instruments, and Systems, 1, 014003

\bibitem[{{Saar} \& {Osten}(1997)}]{SaarOsten97}
{Saar}, S.~H. \& {Osten}, R.~A. 1997, \mnras, 284, 803

\bibitem[{{Samadi} {et~al.}(2007){Samadi}, {Georgobiani}, {Trampedach}, {Goupil}, {Stein}, \& {Nordlund}}]{Samadi07}
{Samadi}, R., {Georgobiani}, D., {Trampedach}, R., {et~al.} 2007, \aap, 463, 297

\bibitem[{{Samadi} {et~al.}(2005){Samadi}, {Goupil}, {Alecian}, {Baudin}, {Georgobiani}, {Trampedach}, {Stein}, \& {Nordlund}}]{Samadi05}
{Samadi}, R., {Goupil}, M.~J., {Alecian}, E., {et~al.} 2005, Journal of Astrophysics and Astronomy, 26, 171

\bibitem[{{Schou}(2018)}]{Schou18}
{Schou}, J. 2018, \aap, 617, A111

\bibitem[{{Spergel} {et~al.}(2015){Spergel}, {Gehrels}, {Baltay}, {Bennett}, {Breckinridge}, {Donahue}, {Dressler}, {Gaudi}, {Greene}, {Guyon}, {Hirata}, {Kalirai}, {Kasdin}, {Macintosh}, {Moos}, {Perlmutter}, {Postman}, {Rauscher}, {Rhodes}, {Wang}, {Weinberg}, {Benford}, {Hudson}, {Jeong}, {Mellier}, {Traub}, {Yamada}, {Capak}, {Colbert}, {Masters}, {Penny}, {Savransky}, {Stern}, {Zimmerman}, {Barry}, {Bartusek}, {Carpenter}, {Cheng}, {Content}, {Dekens}, {Demers}, {Grady}, {Jackson}, {Kuan}, {Kruk}, {Melton}, {Nemati}, {Parvin}, {Poberezhskiy}, {Peddie}, {Ruffa}, {Wallace}, {Whipple}, {Wollack}, \& {Zhao}}]{Roman}
{Spergel}, D., {Gehrels}, N., {Baltay}, C., {et~al.} 2015, arXiv e-prints, arXiv:1503.03757

\bibitem[{{Stassun} {et~al.}(2019){Stassun}, {Oelkers}, {Paegert}, {Torres}, {Pepper}, {De Lee}, {Collins}, {Latham}, {Muirhead}, {Chittidi}, {Rojas-Ayala}, {Fleming}, {Rose}, {Tenenbaum}, {Ting}, {Kane}, {Barclay}, {Bean}, {Brassuer}, {Charbonneau}, {Ge}, {Lissauer}, {Mann}, {McLean}, {Mullally}, {Narita}, {Plavchan}, {Ricker}, {Sasselov}, {Seager}, {Sharma}, {Shiao}, {Sozzetti}, {Stello}, {Vanderspek}, {Wallace}, \& {Winn}}]{TIC}
{Stassun}, K.~G., {Oelkers}, R.~J., {Paegert}, M., {et~al.} 2019, \aj, 158, 138

\bibitem[{{Tassoul}(1980)}]{Tassoul80}
{Tassoul}, M. 1980, \apjs, 43, 469

\bibitem[{{Teixeira} {et~al.}(2009){Teixeira}, {Kjeldsen}, {Bedding}, {Bouchy}, {Christensen-Dalsgaard}, {Cunha}, {Dall}, {Frandsen}, {Karoff}, {Monteiro}, \& {Pijpers}}]{Teixeira09}
{Teixeira}, T.~C., {Kjeldsen}, H., {Bedding}, T.~R., {et~al.} 2009, \aap, 494, 237

\bibitem[{{Vaughan} \& {Preston}(1980)}]{VaughanPreston80}
{Vaughan}, A.~H. \& {Preston}, G.~W. 1980, \pasp, 92, 385

\bibitem[{{Verner} {et~al.}(2011){Verner}, {Elsworth}, {Chaplin}, {Campante}, {Corsaro}, {Gaulme}, {Hekker}, {Huber}, {Karoff}, {Mathur}, {Mosser}, {Appourchaux}, {Ballot}, {Bedding}, {Bonanno}, {Broomhall}, {Garc{\'\i}a}, {Handberg}, {New}, {Stello}, {R{\'e}gulo}, {Roxburgh}, {Salabert}, {White}, {Caldwell}, {Christiansen}, \& {Fanelli}}]{Verner11}
{Verner}, G.~A., {Elsworth}, Y., {Chaplin}, W.~J., {et~al.} 2011, \mnras, 415, 3539

\bibitem[{{Vidotto} {et~al.}(2016){Vidotto}, {Donati}, {Jardine}, {See}, {Petit}, {Boisse}, {Boro Saikia}, {H{\'e}brard}, {Jeffers}, {Marsden}, \& {Morin}}]{Vidotto16}
{Vidotto}, A.~A., {Donati}, J.~F., {Jardine}, M., {et~al.} 2016, \mnras, 455, L52

\bibitem[{{Viswanath} {et~al.}(2021){Viswanath}, {Janson}, {Dahlqvist}, {Petit dit de la Roche}, {Samland}, {Girard}, {Pathak}, {Kasper}, {Feng}, {Meyer}, {Boehle}, {Quanz}, {Jones}, {Absil}, {Brandner}, {Maire}, {Siebenmorgen}, {Sterzik}, \& {Pantin}}]{Viswanath21}
{Viswanath}, G., {Janson}, M., {Dahlqvist}, C.-H., {et~al.} 2021, \aap, 651, A89

\bibitem[{{{\v{S}}ubjak} {et~al.}(2023){{\v{S}}ubjak}, {Lodieu}, {Kab{\'a}th}, {Boffin}, {Nowak}, {Grundahl}, {B{\'e}jar}, {Zapatero Osorio}, \& {Antoci}}]{Subjak23}
{{\v{S}}ubjak}, J., {Lodieu}, N., {Kab{\'a}th}, P., {et~al.} 2023, \aap, 671, A10

\bibitem[{{White} {et~al.}(2012){White}, {Bedding}, {Gruberbauer}, {Benomar}, {Stello}, {Appourchaux}, {Chaplin}, {Christensen-Dalsgaard}, {Elsworth}, {Garc{\'\i}a}, {Hekker}, {Huber}, {Kjeldsen}, {Mosser}, {Kinemuchi}, {Mullally}, \& {Still}}]{White12}
{White}, T.~R., {Bedding}, T.~R., {Gruberbauer}, M., {et~al.} 2012, \apjl, 751, L36

\bibitem[{{White} {et~al.}(2011{\natexlab{a}}){White}, {Bedding}, {Stello}, {Appourchaux}, {Ballot}, {Benomar}, {Bonanno}, {Broomhall}, {Campante}, {Chaplin}, {Christensen-Dalsgaard}, {Corsaro}, {Do{\v{g}}an}, {Elsworth}, {Fletcher}, {Garc{\'\i}a}, {Gaulme}, {Handberg}, {Hekker}, {Huber}, {Karoff}, {Kjeldsen}, {Mathur}, {Mosser}, {Monteiro}, {R{\'e}gulo}, {Salabert}, {Silva Aguirre}, {Thompson}, {Verner}, {Morris}, {Sanderfer}, \& {Seader}}]{White11_1}
{White}, T.~R., {Bedding}, T.~R., {Stello}, D., {et~al.} 2011{\natexlab{a}}, \apjl, 742, L3

\bibitem[{{White} {et~al.}(2011{\natexlab{b}}){White}, {Bedding}, {Stello}, {Christensen-Dalsgaard}, {Huber}, \& {Kjeldsen}}]{White11_2}
{White}, T.~R., {Bedding}, T.~R., {Stello}, D., {et~al.} 2011{\natexlab{b}}, \apj, 743, 161

\bibitem[{{Zhou} {et~al.}(2021){Zhou}, {Nordlander}, {Casagrande}, {Joyce}, {Li}, {Amarsi}, {Reggiani}, \& {Asplund}}]{Zhou21}
{Zhou}, Y., {Nordlander}, T., {Casagrande}, L., {et~al.} 2021, \mnras, 503, 13

\end{thebibliography}

\listofobjects

\begin{appendix}

\section{Radial velocities and associated CCF parameters}\label{sec:RVs}

\begin{figure}[!h]
\centering
\includegraphics[angle=90,width=0.37\textheight,trim=0cm 0.1cm 2cm 0.2cm,clip]{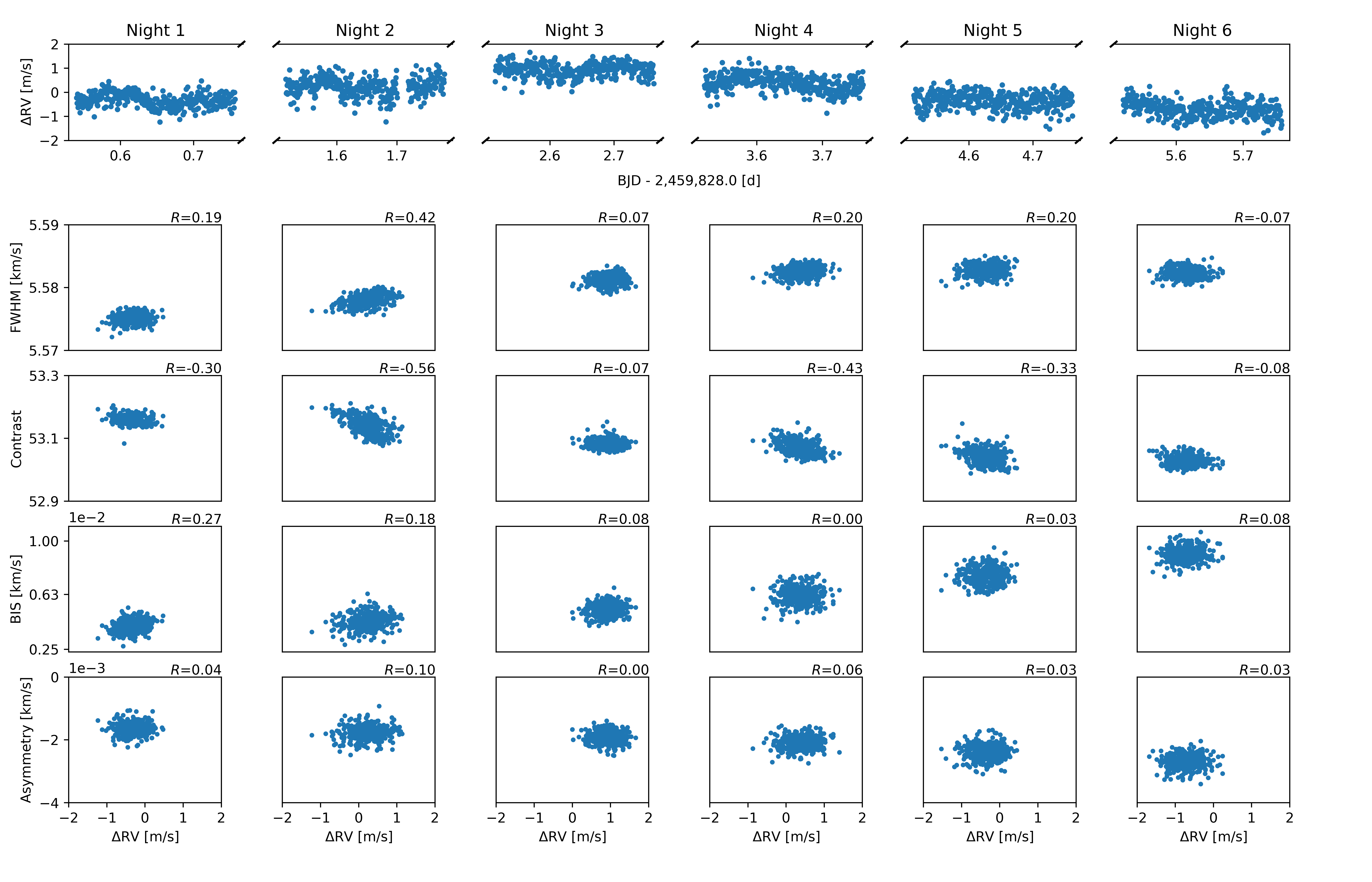}
\caption{Derived radial velocities vs activity proxies. The raw time series (after removal of a constant RV offset) is shown in the leftmost column. A ${\sim} 20$ min gap in the data collection is apparent toward the end of night 2, which resulted from an interruption of the corresponding observation block. The remaining columns show the relationship (on a nightly basis) between the RVs and associated CCF parameters (FWHM, contrast, BIS, and asymmetry). Each panel is supplemented by Pearson's correlation coefficient, $R$.}
\label{fig:ccf_par}
\end{figure}
\FloatBarrier

\noindent Spectra were reduced using the ESPRESSO DRS (version 3.0.0). We searched for correlations between the RVs and a number of activity proxies derived concurrently from the CCFs, namely the full width at half maximum (FWHM), contrast, bisector span (BIS), and asymmetry. From Fig.~\ref{fig:ccf_par}, no obvious correlations can be seen (also supported by the correlation coefficients displayed in each panel). A trend in the different CCF metrics is nevertheless evident throughout the observing run.

\section{Power spectrum prewhitening}\label{sec:Prewhitening}

\begin{figure}[!t]
\centering
\includegraphics[width=0.5\textwidth,trim=3.25cm 7.2cm 3.4cm 1.45cm,clip]{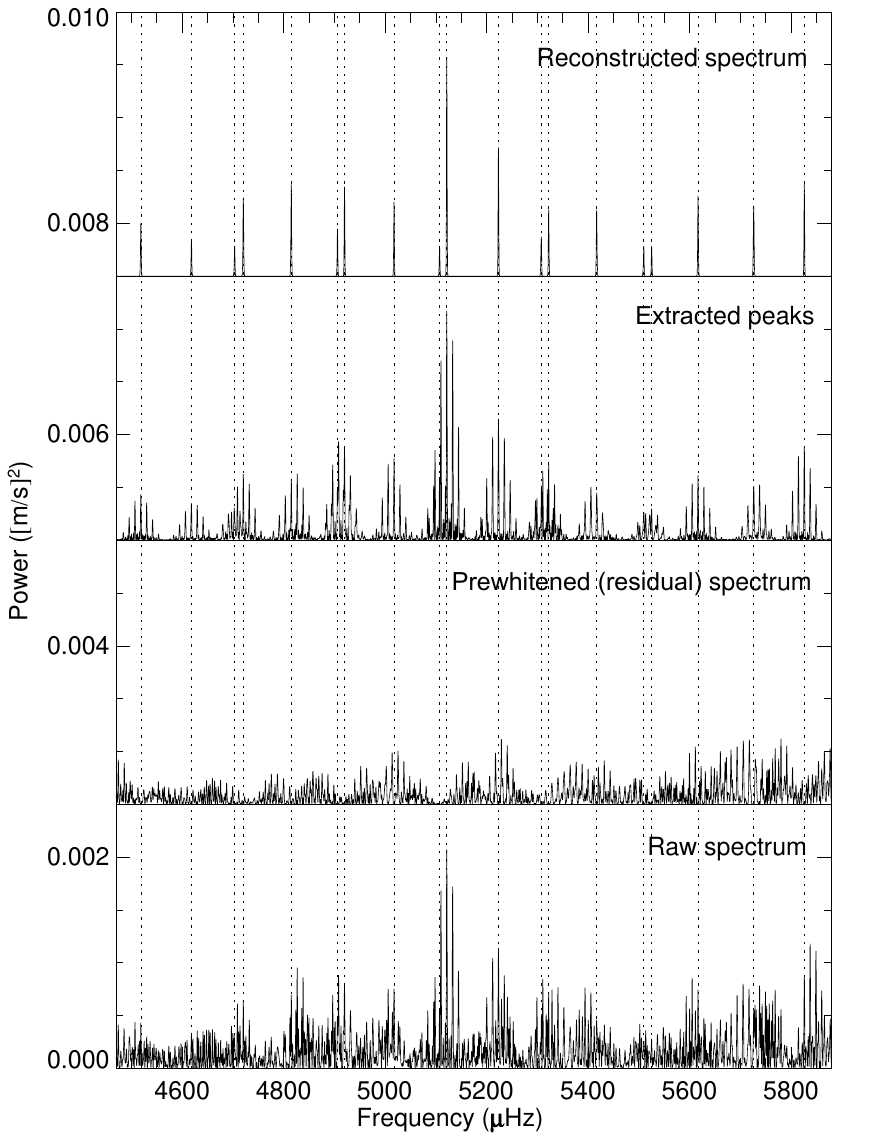}
\caption{Power spectrum prewhitening. The outcome of having extracted all 19 identified modes is illustrated (see text for details).}
\label{fig:prewhitening}
\end{figure}

The outcome of the iterative prewhitening procedure (after extraction of all 19 identified modes) is illustrated in Fig.~\ref{fig:prewhitening}. The bottom panel shows the raw power spectrum (i.e. before prewhitening). The two panels immediately above the bottom panel display the prewhitened (or residual) power spectrum and the extracted modes (along with their aliases). The top panel shows a reconstructed power spectrum where the effect of the spectral window has been deconvolved.

\FloatBarrier

\section{Chromospheric emission and magnetic activity cycle}\label{sec:Activity}

Characterised by a median chromospheric emission level of $\log R^{\prime}_{\rm HK}\!=\!-4.75\:{\rm dex}$ \citep[][]{GomesdaSilva21}, $\epsilon$~Indi is more active than the Sun \citep[$\log R^{\prime}_{{\rm HK},\sun}\!=\!-4.91\:{\rm dex}$;][]{MamajekHillenbrand08}. It lies close to the Vaughan--Preston gap \citep[][]{VaughanPreston80}, identified as a dearth of stars in the distribution of $\log R^{\prime}_{\rm HK}$ for main-sequence stars at about $-4.75\:{\rm dex}$. Moreover, $\epsilon$ Indi has an activity dispersion of $\log\sigma(R_5)\!=\!-0.84\:{\rm dex}$, where $R_5\!=\!R^{\prime}_{\rm HK} \times 10^5$, hence displaying a typical activity variability for a K dwarf \citep[][]{GomesdaSilva21}.

\begin{figure}[!t]
\centering
\includegraphics[width=0.5\textwidth,trim=0.6cm 0.5cm 0.3cm 0.3cm,clip]{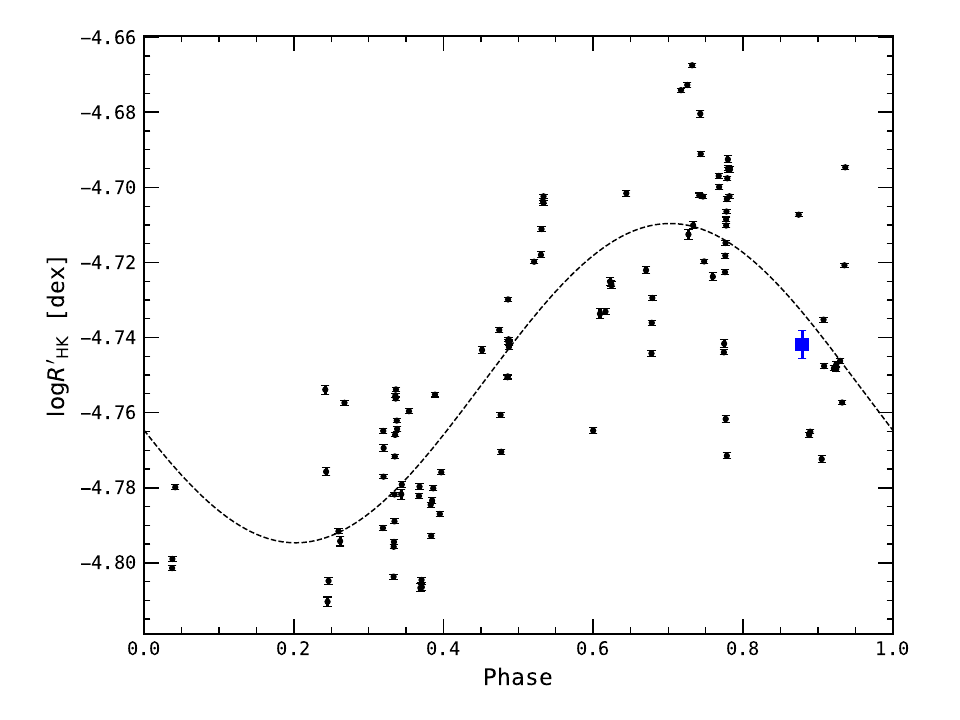}
\caption{Activity cycle of $\epsilon$ Indi as traced by $\log R^{\prime}_{\rm HK}$. The long-term variation of $\log R^{\prime}_{\rm HK}$ is phase-folded onto the estimated cycle period of ${\sim} 2600$ days and modelled assuming a simple sinusoidal behaviour (dashed curve). The black dots represent HARPS observations, while the blue square represents the single-epoch, mean ESPRESSO emission level. The error bar on the latter is given by the standard deviation of the observations over the ESPRESSO campaign.
}
\label{fig:cycle}
\end{figure}

Based on 4293 HARPS archival observations obtained between 2003 and 2016, binned to 112 nights, we extracted the \ion{Ca}{ii} H\&K index using \texttt{ACTIN~2} \citep[][]{ACTIN,GomesdaSilva21} and converted it to $\log R^{\prime}_{\rm HK}$ following \citet[][]{GomesdaSilva21}. We next modelled the activity cycle of $\epsilon$~Indi, as traced by $\log R^{\prime}_{\rm HK}$, assuming a simple sinusoidal behaviour. Figure~\ref{fig:cycle} shows the long-term variation of $\log R^{\prime}_{\rm HK}$ phase-folded onto the estimated (from a periodogram analysis) cycle period, $P_\mathrm{cyc}$, of ${\sim} 2600$ days (or ${\sim} 7$ yr). The scatter of the HARPS observations about the activity cycle model is caused by rotational activity variations. We note that a period of about 2500 days for the primary magnetic cycle is reported by \citet{Feng19}, in good agreement with the value estimated here.

We used the ESPRESSO spectra obtained herein to compute $\log R^{\prime}_{\rm HK}$ following the same procedure as above, having determined a mean chromospheric emission level of  $\log R^{\prime}_{\rm HK}\!=\!-4.742\pm0.004\:{\rm dex}$ (blue square in Fig.~\ref{fig:cycle}). The ESPRESSO emission level is very close to the value predicted by our simple model. Moreover, it is apparent that the ESPRESSO observations were obtained during the descending phase of the cycle, following the last maximum of activity in 2021.

Based on the activity cycle period, $P_\mathrm{cyc}$, estimated above and the rotation period estimate, $P_\mathrm{rot}$, from \citet{Feng19}, we evaluated the corresponding rates as $\omega_\mathrm{cyc}\!\equiv\!2\pi/P_\mathrm{cyc}$ and $\Omega\!\equiv\!2\pi/P_\mathrm{rot}$, yielding $\log \omega_\mathrm{cyc} \! \simeq \! -2.62$ and $\log \Omega \! \simeq \! -0.76$. When placed in the context of the correlation between activity cycle and rotation rates recently analysed by \citet{Bonanno22} for an extended Mt.~Wilson sample of 67 stars, $\epsilon$~Indi appears to belong to what is called the upper branch, consisting of stars for which $\omega_\mathrm{cyc}$ increases with increasing $\Omega$. According to the same study, stars belonging to the upper branch tend to be more metal-poor than stars in the lower branch (to which the Sun belongs), in line with the subsolar metallicity of $\epsilon$~Indi. $\epsilon$~Indi is thus likely characterised by an $\alpha\Omega$ type of dynamo action in which global stellar rotation plays a primary role in setting the efficiency of the dynamo process since, according to mixing-length theory, a reduced metallicity enhances the eddy diffusivity of the plasma in the outer convection zone.

\FloatBarrier

\section{Revisiting HD~40307}\label{sec:HD40307}

With the goal of reassessing the tentative claim of p-mode detection for HD~40307 \citep{ESPRESSO}, we re-reduced the 2018 ESPRESSO GTO data using the latest version (3.0.0) of the DRS, having computed a noise-optimised power spectrum in the same manner as described in Sect.~\ref{sec:Pow}. We next fitted a background profile plus a Gaussian envelope to the power spectrum using the high-DImensional And multi-MOdal NesteD Sampling \citep[\texttt{DIAMONDS};][]{DIAMONDS} Bayesian software. The Gaussian envelope is used to describe a possible power excess due to solar-like oscillations and was centred at different proxies for $\nu_{\rm max}$ in the range 3--$6\:{\rm mHz}$. Finally, we computed the Bayes factor ($B$) in favour of a model containing the envelope (over a model without the envelope) to test the statistical significance of the presence of a power excess. The resulting logarithmic factor of $\ln B \! < \! -2$ (irrespective of the $\nu_{\rm max}$ proxy adopted) provides strong evidence \citep{Jeffreys61} against the detection of p modes, at odds with the findings of \citet{ESPRESSO}.

The average photon-noise level in the amplitude spectrum, as measured in the range\footnote{Given the relatively low Nyquist frequency of $6.4\:{\rm mHz}$ (based on a median cadence of $78\:{\rm s}$), we instead measured the photon-noise level in a frequency range below the expected location of the p modes.} 1.5--$3\:{\rm mHz}$, is $4.93\:{\rm cm\,s^{-1}}$, which we adopt as an upper limit on the mode amplitude. The factor of ${\sim} 5$ difference in the photon-noise level relative to that measured for $\epsilon$~Indi ($0.94\:{\rm cm\,s^{-1}}$) is mostly accounted for by the nearly ten times lower brightness of HD~40307 and shorter effective length of its observing campaign (1150 spectra with a fixed exposure time of $30\:{\rm s}$). We further note that the blue detector of ESPRESSO was prone to, at the time the observations of HD~40307 were collected, a known RV systematic effect\footnote{\url{https://www.eso.org/sci/facilities/paranal/instruments/espresso/ESPRESSO_User_Manual_P109_v2.pdf}} with periodicities of five and seven minutes caused by a temperature instability in the blue cryostat, which may also be contributing to the overall noise budget in this frequency range.

\section{Mode lifetime calibration}\label{sec:Lifetimes_calibration}

\begin{figure}[!t]
\centering
  \subfigure{ \includegraphics[width=.48\textwidth]{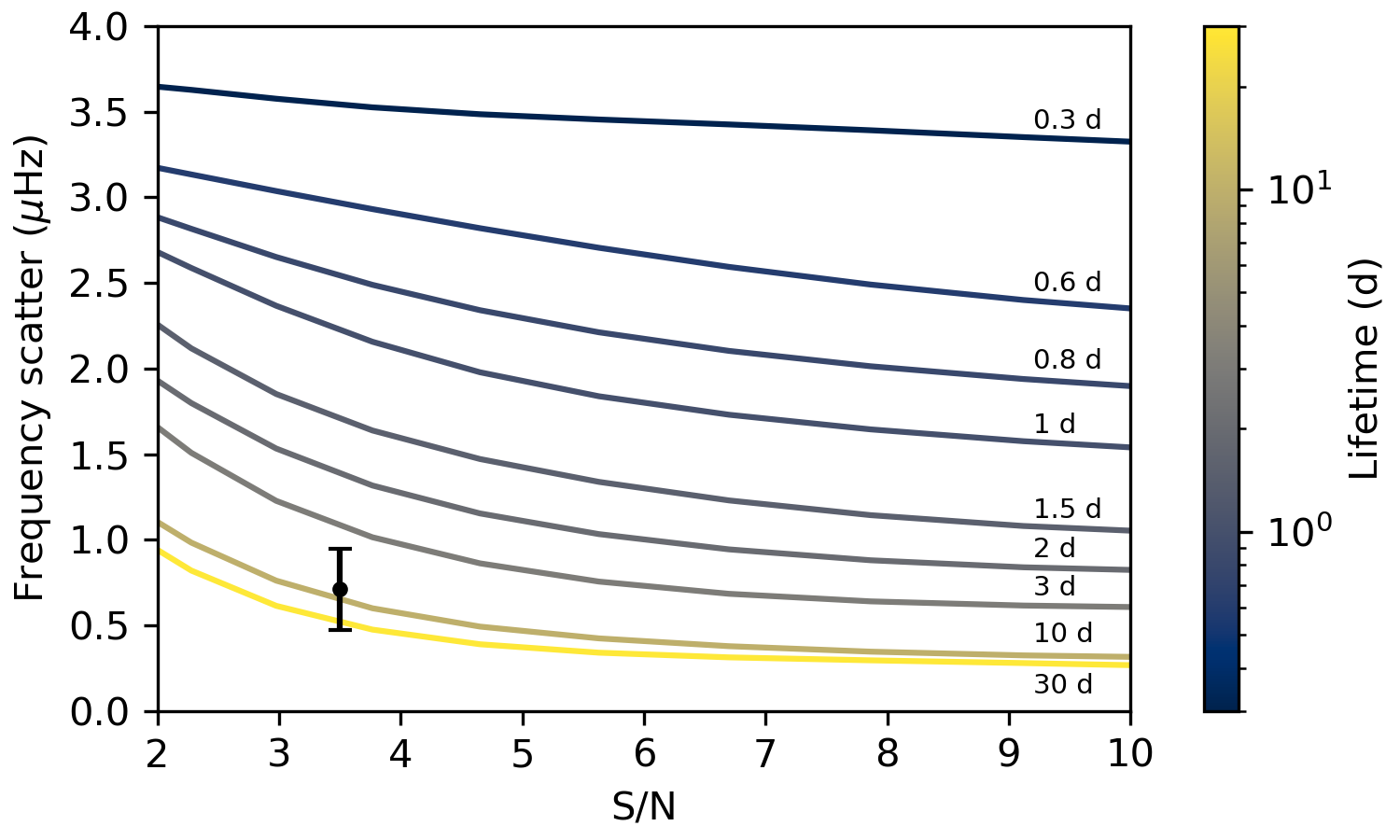}}
  \subfigure{ \includegraphics[width=.48\textwidth]{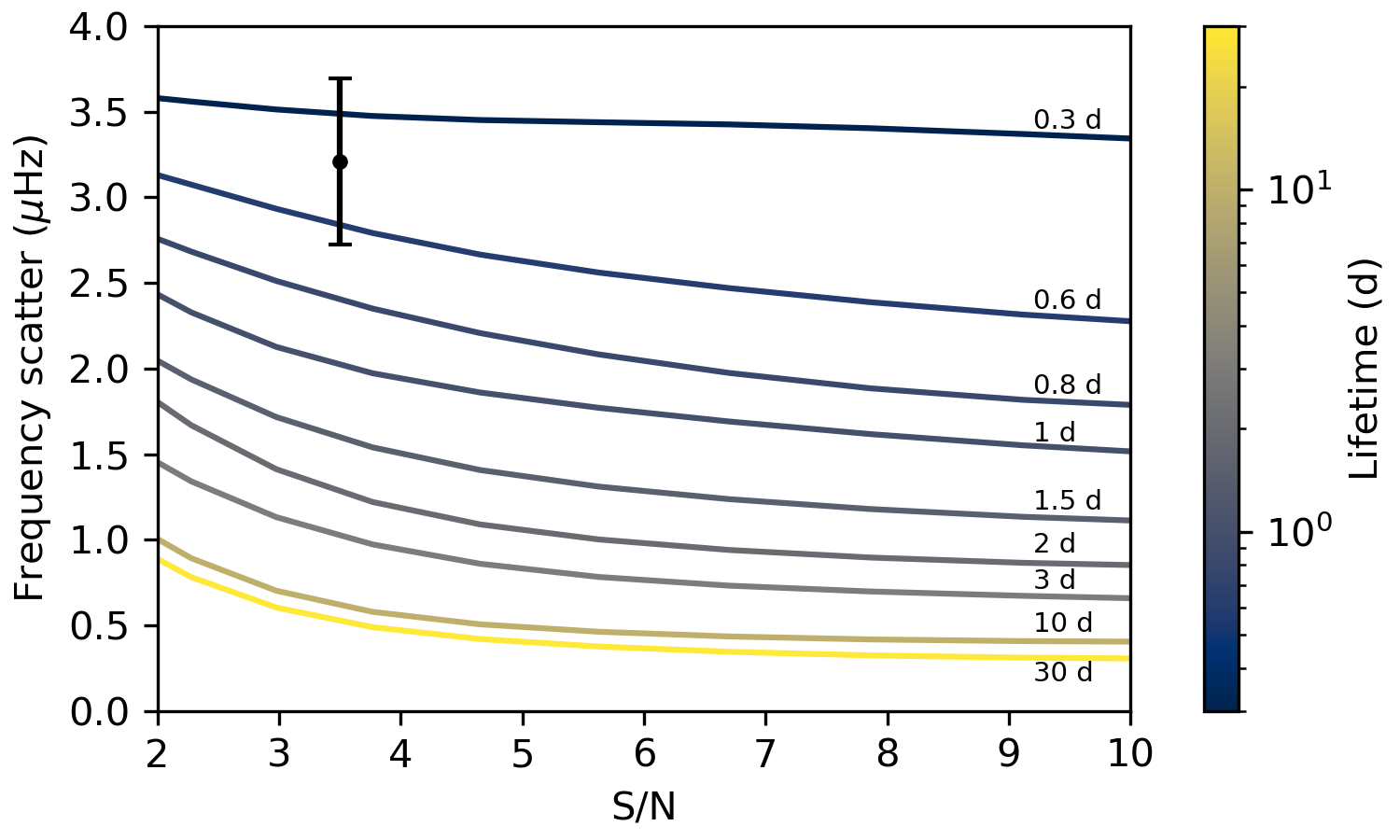}}   
  \caption{Mode lifetime calibration (top: radial modes; bottom: dipole modes) for the $\epsilon$~Indi observing window. The solid lines are the result of simulations (see text for details) and show frequency scatter vs $S/N$ for a range of input mode lifetimes. The measured frequency scatter is represented by a black symbol in both panels, placed at a fiducial $S/N\!=\!3.5$ level (corresponding to the typical $S/N$ of radial modes near $\nu_{\rm max}$).}\label{fig:lifetime}
\end{figure}

We conducted simulations following \citet{Kjeldsen05}, having made use of \texttt{AADG3} \citep[][]{Ball18} to generate artificial time series. Each simulation contains a single input frequency of varying mode lifetime (0.3, 0.6, 0.8, 1, 1.5, 2, 3, 10, and 30 days) and $S/N$, and was sampled adopting the $\epsilon$~Indi observing window. Simulations were run considering both radial (which are not impacted by rotation) and dipole modes. The $\ell\!=\!1$ simulations also included rotational splittings assuming an edge-on configuration\footnote{An edge-on configuration (i.e. a stellar inclination angle $i\!=\!90\degree$) is assumed here for illustrative purposes only, as it maximises the relative visibility of the two sectoral modes in a dipole triplet. We note that the best orbital solution of \citet[][]{Feng19} is instead characterised by $i\!=\! 64.25{\degree}^{+13.80}_{-6.09}$.} and a rotation period\footnote{The rotation period estimate adopted throughout this work, $P_{\rm rot}\!=\!35.732^{+0.006}_{-0.003}\:{\rm d}$, is from \citet[][]{Feng19}. The authors argue that this estimate, which is derived from a relatively large data set of high-precision RVs and multiple activity indicators, is a more reliable estimate than the 22-day rotation period quoted by \citet[][]{SaarOsten97}. A 37.2-day estimate based on an unpublished ZDI analysis is quoted by \citet[][]{Vidotto16}.} of 36 days. Mode lifetimes may then be inferred, for a given fiducial $S/N$ level, by comparing\footnote{The simulated and observed frequency scatter are given by the median absolute deviation.} the simulated frequency scatter with the scatter of the observed frequencies about the asymptotic values given by Eq.~(\ref{eq:asymptotic}). Figure~\ref{fig:lifetime} shows the outcome of this calibration procedure for radial (top panel) and dipole (bottom panel) modes.

We explored possible physical causes for the enhanced scatter about the $\ell\!=\!1$ power ridge (apparent in the top panel of Fig.~\ref{fig:echelle} and made more evident in the bottom panel of Fig.~\ref{fig:lifetime}), a feature that we find not to be reproduced by model frequencies; the same models rule out undetected $\ell\!=\!3$ modes as the cause since they are expected to lie farther away from the ridge than implied by the observed scatter. Based on the simulation results, the observed scatter of dipole modes is too large to be explained by a combination of mode damping and rotational splittings. If, however, a strong dipole magnetic field is present, this could in principle lead to extra damping of the $\ell\!=\!1$ modes (with respect to the $\ell\!=\!0$ modes). The unpublished, single-epoch Zeeman--Doppler imaging (ZDI) map from \citet{Vidotto16} is indeed dominated by a strong dipole field ($56\%$ of the field energy is in the dipole component, according to their Table 2). A similar field, if concurrent with the ESPRESSO observations, could partially account for the enhanced scatter about the $\ell\!=\!1$ ridge.

\end{appendix}

\end{document}